\documentclass[a4paper, 11pt]{elsarticle} 
\usepackage{hyperref,indentfirst,caption,float,amsmath,tabularx,algorithm,setspace,longtable,multirow, graphicx, booktabs, array, caption,enumitem,cleveref,color,soul,lscape, appendix, amssymb}

\allowdisplaybreaks
 
\usepackage[margin=1in]{geometry}
\usepackage{url}
\usepackage{algorithm,algpseudocode}

\newcolumntype{P}[1]{>{\centering\arraybackslash}p{#1}}
\newcolumntype{M}[1]{>{\centering\arraybackslash}m{#1}}
\makeatletter
\def\BState{\State\hskip-\ALG@thistlm}
\makeatother

\journal{}
\biboptions{numbers,sort&compress}

\DeclareMathOperator*{\argmax}{argmax}

\usepackage{hanging}
\usepackage{tikz}
\usetikzlibrary{positioning}

\usetikzlibrary{calc,positioning,shapes.geometric, arrows}
\tikzstyle{arrow} = [thick,->,>=stealth]

\tikzset{
	every node/.style={draw=black, text centered,text width=15cm},
}

\usepackage{titlesec}
\titleformat{\subsection}
{\bfseries\itshape}{\thesubsection}{1em}{}

\begin{document}
	
\begin{frontmatter}
%\title{Uncovering Business Intelligence from Online Airline Reviews: An Unsupervised Text Analytics Approach }

%\title{Aspect-based opinion analysis of online reviews for uncovering business intelligence in the airline industry: An Unsupervised Text Analytics Approach }

\title{Discovering Airline-Specific Business Intelligence from Online Passenger Reviews: An Unsupervised Text Analytics Approach}

\author[1,2]{Sharan Srinivas\corref{cor1}}
\ead{SrinivasSh@missouri.edu}

\author[3]{Surya Ramachandiran}

\cortext[cor1]{Corresponding author}

\address[1]{Department of Industrial and Manufacturing Systems Engineering, University of Missouri, United States}

\address[2]{Department of Marketing, University of Missouri, United States}

\address[3]{Department of Mechanical Engineering, National Institute of Technology Tiruchirappalli, India}

\begin{abstract}
	\setstretch{1.15}
	Driven by the low passenger satisfaction and fierce competition in the airline industry, carriers seek to  deliver a flawless  travel experience as well as exceed passenger expectations to attract and retain customers. To understand the important dimensions of service quality from the passenger’s perspective and tailor service offerings for competitive advantage, airlines can capitalize on the abundantly available online customer reviews (OCR). The objective of this paper is to discover company- and competitor-specific intelligence from OCR using an unsupervised text analytics approach. First, the key aspects (or topics) discussed in the OCR are extracted using three topic models - probabilistic latent semantic analysis (pLSA) and two variants of Latent Dirichlet allocation (LDA-VI and LDA-GS). Subsequently, we propose an ensemble-assisted topic model (EA-TM), which integrates the individual topic models, to classify each review sentence to the most representative aspect. Likewise, to determine the sentiment (positive, neutral and negative) corresponding to a review sentence, an ensemble sentiment analyzer (E-SA), which combines the predictions of three opinion mining methods (AFINN, SentiStrength and VADER), is developed. An aspect-based opinion summary (AOS), which provides a snapshot of passenger-perceived strengths and weaknesses of an airline, is established by consolidating the sentiments associated with each aspect. Furthermore, a bi-gram analysis of the labeled OCR is employed to perform root cause analysis within each identified aspect. A case study involving 99,147 airline reviews of a US-based target carrier and four of its competitors is used to validate the proposed approach. The results indicate that a cost- and time-effective performance summary of an airline and its competitors can be obtained from OCR. Moreover, the proposed EA-TM and E-SA achieved 17-23\% and 9-20\% higher average accuracy over individual benchmark models, respectively. Finally, besides providing theoretical and managerial implications based on our results, we also provide implications for post-pandemic preparedness in the airline industry considering the unprecedented impact of coronavirus disease 2019 (COVID-19) and  predictions on  similar pandemics in the future.
\medskip
\end{abstract}

\begin{keyword}
Airline industry \sep text mining \sep online customer reviews \sep topic modeling \sep sentiment analysis \sep business intelligence \sep competitive intelligence.
\end{keyword}
\end{frontmatter}

\newpage
%\linenumbers

\section{Introduction}
The competition in the airline industry has become fierce due to deregulation, and the proliferation of low-cost carriers (LCC).  To sustain in this challenging environment and gain a competitive advantage, airline companies are striving for passenger satisfaction and loyalty. They offer competitive fares and provide differentiated services, such as priority check-in and on-board internet, to attract and retain customers. Nevertheless, the airline industry is frequently ranked low in customer satisfaction \cite{Mazzeo2003}. Further, the U.S. Congress' consistent efforts and the consequent establishment of the "Airline Customer Service Commitment" too failed to ameliorate passenger dissatisfaction \cite{Mazzeo2003}. Improving in critical areas of service quality is essential for customer contentment as well as for enhancing business performance and customer loyalty \cite{Sezgen2019}. Specifically, identifying service quality attributes that are most-valued by passengers and meeting/surpassing their expectation for those attributes is crucial for achieving high customer satisfaction \cite{Sezgen2019}. 

Traditionally, survey studies were used to understand customer expectations and satisfaction in the airline industry. However, Lucini et al. \cite{Lucini2020} criticized such survey methods as the aspects focused tend to be based on the knowledge of management or researchers and may not mirror the customer’s actual needs, resulting in an inconsistent measurement of perceived service quality. Another drawback to these techniques is that they limit the response a customer can provide, and the expected structured response also distorts the customer’s intrinsic voice \cite{Korfiatis2019}. Finally, many of these survey forms are completed perfunctorily, and some customers even respond randomly, leading to a fictitious understanding of customer perception, which is more harmful to any business \cite{Lucini2020}.
%These hearings were followed by the airlines executing  an Airline Customer Service Commitment, which led to more hearings, but no significant improvement in customer satisfaction (U.S. Department of Transportation, 2007). 
%The services offered by an airline depends on its business model. Specifically, airlines across the globe can be categorized into two groups for short-haul services: full service carriers (FSC) and low-cost carriers (LCC). Th LCCs  offer  economical airfares in the exchange for the elimination of other services such as service classes, in-flight meals, and entertainment as they are solely focused on minimizing operating costs. On the other hand, FSC are companies with large scale complex operations that offer better services (e.g., meals, in-flight entertainment, multiple classes) and comfort. The notion of service quality has increased with competition in airline operations. With some LCC outgrowing FSC by over 100\% since 2011, the latter must find other ways to differentiate themselves (Kasper 2017).

With increasing accessibility to Web 2.0 technologies and expansion of social media websites, more customers express their opinions/experiences about a product or service on online forums without much restrictions \cite{Rajendran2020}.  Even though online customer reviews (OCR) are subjective and represent a subset of actual customers, they are considered  highly trustworthy for understanding a customer’s perspective \cite{Srinivas2019}. Besides, OCR also influences the decisions made by potential customers \cite{Salehan2016}. A negative OCR can undermine the company's reputation and also spreads five to six times quicker than positive reviews \cite{Salehan2016}. Thus, businesses are faced with the need to understand and analyze these unstructured review data. Further, these reviews have become widely available and easily accessible, allowing businesses to effortlessly understand their strengths and weaknesses from a customer's perspective \cite{Salehan2016}. Hence it is presumed that the airline industry can also use OCR to improve its passenger experience. Typically, a customer review discusses several aspects of a product/service and has a sentiment associated with each aspect. For instance, consider the passenger's review given in Figure \ref{fig:Fig1}. The first two sentences express a negative sentiment towards on-time performance and seat comfort, respectively. Whereas, the third sentence portrays the flight crew in a positive manner. Thus, each sentence in this review can be classified into a topic and its corresponding sentiment. A topic-wise summary of the customer opinion (i.e., proportion of positive, negative and neutral reviews for each topic) can be obtained by aspect and sentiment labeling of representative review sentences.  

\begin{figure}[H]

	\begin{tikzpicture}[mybox/.style={minimum width=4cm,draw,thick,align=center,minimum height=1.8cm}]
	\node[mybox] () {\emph{``...the flight was delayed by 2.5 hours due to mechanical issues. Besides, the economy seats were small and cramped. But the crew and pilots were amazing.""}};
	\end{tikzpicture}\caption{Sample airline review by a passenger}\label{fig:Fig1}

\end{figure}

Since OCR is a vast collection of unstructured text, processing them to gain insights is a challenge. However, with advances in natural language processing (NLP), many approaches have been introduced to effectively mine unstructured text \cite{Likhitha2019, Ribeiro2016}. Hence, businesses can process millions of these unstructured textual data and gain insights from them without having to read or intervene manually. Prior research has also considered text analytics of airline reviews to identify the key dimensions impacting customer satisfaction \cite{Korfiatis2019,Sezgen2019,Lucini2020}. While these works extract the topics discussed in OCR for a group of airlines, they do not provide a topic-wise sentiment summary of each airline, which is necessary to investigate the strengths and weaknesses of an airline and its competitors. Besides, these studies do not examine the prominent underlying cause of a positive or negative reviews.  

This paper aims to (i) identify the key aspects/topics affecting passenger satisfaction based on thousands of OCR, (ii) automatically classify each review sentence to its most prominent topic and sentiment (iii) provide a snapshot of topic-wise opinion summary of an airline (iv) extract business and competitive intelligence from OCR. We integrate three different techniques, namely, topic modeling, sentiment analysis and root cause identification, to achieve the aforementioned goals. First, an unsupervised ensemble technique based on topic modeling is proposed for aspect identification and labeling of review sentences. Likewise, an ensemble opinion analyzer is proposed to automatically detect the sentiment of each review sentence. Subsequently, these results are combined to obtain the sentiment summary for each topic. Besides, a data-driven root cause analysis is performed to identify the main reasons for customer satisfaction or dissatisfaction towards each aspect. Finally, the insights obtained are leveraged to provide managerial implications. Since the proposed approach provides comprehensive insights by automatically analyzing passenger reviews, it is efficient in terms of cost and time. Moreover, it serves as a decision support system to monitor the changes in customer preferences and enable the airline management establish their priorities, which in turn can improve passenger satisfaction, airline reputation, customer loyalty, and competitive advantage.  

The remainder of the paper is organized as follows. A review of other relevant studies in the literature is presented in Section 2. Section 3 provides a detailed description of the proposed approach. The case study descriptions, along with the results, are discussed in Section 4. Lastly, conclusions and the scope for future research are given in Section 5.

\section{Literature Review}
Capturing the customers' voice is crucial to understand their expectations and perceptions needed to bridge service quality gaps and satisfy customers \cite{Lucini2020}. Since this research focuses on extracting insights based on the voice of passengers, we review some relevant notable works in this section. Subsequently, we identify the gaps in literature and delineate the contributions of this research.

\subsection{Measuring Passenger's Expectation and Perception}
Conventionally, the customer's perspective of a service is measured using specifically-designed survey instruments \cite{Parasuraman1988,Cronin1992}. Notably, SERVQUAL, a popular multi-item survey instrument developed to quantify the customer's expectations and perception pertaining to the five dimensions (reliability, assurance, tangibles, empathy, responsiveness) of service quality, is majorly employed. Moreover, to specifically capture the voice of airline passengers, Bari et al. \cite{Bari2001} proposed a modified questionnaire, AIRQUAL, by considering the following personalized dimensions that are specific to air travel - airline tangibles, terminal tangibles, personnel, empathy, and image.

These survey instruments have been widely used to understand the passenger's perceptions and expectations in the airline industry. Pakdil and Aydin \cite{Pakdil2007} developed a 35-item questionnaire based on SERVQUAL dimensions to assess the service quality of a Turkish airline. Their results indicated responsiveness (e.g., efficient check-in, willingness to help) to be a top-priority for passengers, while availability (e.g., presence of global alliance and travel-related partner) was perceived to be an non-essential element of service. Koklic et al. \cite{Koklic2017} leveraged the AIRQUAL scale to study the relationship between airline tangibles, service quality, customer satisfaction, and the intention to recommend and repurchase. They found increased service quality to be positively associated with customer satisfaction, which in turn influenced the intent to repurchase and recommend. Instead of relying on existing instruments, Kurtulmuşoğlu et al. \cite{Kurtulmusoglu2016} developed a customized survey to capture the voice of airline customers to specifically understand their priorities when selecting a preferred airline. Their study revealed low fares, high reliability (or punctuality), and online booking facility to be the top priorities for customers, while in-flight meal service and reward programs were found to least impact the passenger's preference in airlines.

Though most of the aforementioned researchers utilized surveys/questionnaires to capture their customers' opinions, these methods have limitations stemming from their restricted and predefined nature as enumerated in the earlier section. Owing to these limitations, recent research has capitalized on freely written OCR for capturing customer's intrinsic voice and understanding their expectations and perceptions. While it is difficult to extract information from OCR, it is crucial as it also serves as a link between prior and prospective customers, influencing purchasing decisions \cite{Salehan2016}.

\subsection{Harnessing Unstructured User-Generated Content to Gather Insights}

Many studies have shown the importance of user-generated content (UGC), such as online reviews and tweets, for improving customer satisfaction \cite{Bastani2019}, service quality \cite{Korfiatis2019} , brand reputation \cite{Seo2020}, and loyalty \cite{Liu2019}. One of the frequently used approaches for processing unstructured textual data is topic modeling, an unsupervised machine learning technique capable of scanning a set of documents to identify and extract the hidden semantics that occurs in it \cite{Likhitha2019}. Different variants of topic models are employed to detect dominant topics (or aspects) discussed in customer reviews, tweets, and other textual data from versatile domains \cite{Likhitha2019}. Latent semantic analysis (LSA) is one of the first algorithms for topic modeling, which generates a high-dimensional co-occurrence matrix (or document-term matrix) based on the corpus, and then employs singular value decomposition to decompose them into topics \cite{Deerwester1990}. Nevertheless, LSA is not a generative model and uses many ad-hoc parameters. To overcome these limitations, probabilistic latent semantic analysis (pLSA) was proposed \cite{Hofmann2001}. Subsequently, other unsupervised topic modeling algorithms, such as latent Dirichlet allocation (LDA) \cite{Blei2003}, were proposed to extract topics using different statistical approaches. Another NLP technique to understand customer perception is sentiment analysis (SA) or opinion mining, which  aims to identify and extract subjective information in texts such as opinions, attitudes, and emotions \cite{Ribeiro2016}. The SA methods used to determine the sentiment of a sentence/text can be broadly classified into three types – lexicon-based (uses set of rules and a dictionary of words tagged to their corresponding sentiment polarity), machine learning-based (uses supervised algorithms to learn patterns from sentiment-tagged sentences) and hybrid approach (adopts a combination of lexical features and machine learning algorithms) \cite{Ribeiro2016}.

Topic models have been used to identify the key dimensions of service quality discussed in OCR across several industries \cite{Lucini2020}. Additionally, sentiment analyzers have been employed to detect emotions and opinions associated with the reviews \cite{Ribeiro2016}. Nevertheless, only few studies have employed these techniques to understand drivers of passenger satisfaction in the airline industry \cite{Lucini2020}. Korfiatis et al. \cite{Korfiatis2019} mined multiple airlines' online reviews and extracted 20 aspects of service quality using the structural topic model. Besides, they found good customer service to have the greatest positive impact on the customer rating, while delays and cancellations led to stronger negative impacts. Likewise, Sezgen et al. \cite{Sezgen2019} scraped 5,120 online reviews and extracted key factors affecting passenger satisfaction and dissatisfaction using LSA. Their analysis revealed ten and twelve factors associated with positive and negative reviews, respectively. Specifically, friendliness of staff and ticket prices were found to be critical drivers of satisfaction, while legroom and flight disruptions were the main reasons for dissatisfaction. Recently, Lucini et al. \cite{Lucini2020} identified topics affecting passenger satisfaction using the LDA model and predicted a customer's recommendation  using the aspects as features. 

Few works on airline review analysis have adopted a different approach to detect both topics and the associated sentiments. Instead of using topic models, Siering et al. \cite{Siering2018} used domain-specific words to detect service aspects discussed by the passengers, and lexicon-based approach to determine the sentiment polarity. Further, they leveraged the aspect-level sentiments and machine learning algorithms to predict a customer's recommendation (yes or no) for an airline. Ma et al. \cite{Ma2019} extracted prevalent topics and sentiments from customer tweets following an overbooking crisis encountered by an airline. Their analysis indicated the overall customer perception to be negative towards the incident, and vary over time depending on the crisis management approach adopted by the airline.

Besides understanding the customer's perspectives, OCR is also used to gather business intelligence (BI), which involves the identification of both company-specific and competitive information to aid and improve the decision-making process \cite{Negash2008}. Traditionally, BI is majorly engendered from structured and semi-structured data, but with the increase in availability of web data and the development of tools to mine them, unstructured data can be effectively analyzed to gain intelligence \cite{Likhitha2019, Ribeiro2016}. This has been the case with the airlines' industry as well. Xu et al. \cite{Xu2019} mined online reviews of airlines with service failures to identify causes and therefore implement a recovery strategy to improve the airline's reputation, finance, and customer retention \cite{Xu2019}. Similarly, Hong et al. \cite{Hong2019} analyzed online reviews to identify the marketing performance of individual airlines. Instead of relying on OCR, Hu et al. \cite{Hu2017} utilized tweets pertaining to global airline companies to help them understand their brand perceptions, which is crucial to strategic decision-making. While most of these research works focused on analyzing UGC to help individual companies identify their strengths and weaknesses, very few studies also focused on mining the competitor's UGC to assess potential opportunities and threats (e.g., \cite{Lacic2016,Korfiatis2019}).

\subsection{Research Gaps and Contributions}
Based on the review of the relevant literature, the following gaps have been identified.
\begin{itemize}
	\item Very few works have leveraged text analytics approach for extracting business intelligence from OCR in the airline industry \cite{Siering2018,Korfiatis2019,Sezgen2019,Lucini2020}. 
	\item Due to the lack of competitor information on the public domain, not enough studies have focused on competitive intelligence \cite{Xu2011}, especially in the airline industry.
	\item To the best of our knowledge, none of the prior studies have used an automated approach for classifying/labeling a review sentence into a topic. Most of them have relied on supervised learning approaches for sentence classification.  The major limitation of such an approach is that they require large samples of labeled datasets. Therefore, if domain-specific labeled data is not readily available, such a method is time-consuming as it requires manual annotation of each review sentence. 
	\item None of the extant studies have implemented unsupervised generation of topic-wise sentiment summary. As the customer's sentiment towards each aspect may change over time, adopting an automated approach will provide an efficient and economic summary of customer perception as opposed to a manual/supervised approach.  
	\item Existing studies have analyzed airline reviews to extract key topics/aspects, but have not explored the underlying cause of customer satisfaction or dissatisfaction within each aspect. Such root cause analysis is crucial to gather insights on customer perception and generating actionable insights \cite{Rajendran2020}.

\end{itemize}

This research seeks to address the aforementioned gaps in the literature pertaining to airline review analysis. Specifically, the OCR of a target airline and its competitors are analyzed to obtain business intelligence including competitor information. We propose a novel unsupervised ensemble topic model and sentiment analyzer to automatically classify each review sentence into one of the identified topics and sentiments. Previous research in text mining has demonstrated the discriminative capability of topic models to automatically label sentences \cite{Gong2018}. However, such an approach might not provide good classification accuracy. Since extant research has indicated that an ensemble approach can outperform single classification models \cite{Srinivas2019}, we adopt a similar strategy for topic and sentiment classification of each review sentence. Upon validation, the proposed models are employed to obtain an aspect-based opinion summary for each airline under consideration to enable the managers of the target airline better understand their strengths, weaknesses, opportunities, and threats. Besides, to facilitate data-driven actionable insights, the underlying cause for an overall positive or negative sentiment towards a topic is analyzed for the target airline. Typically, a root cause analysis is conducted by manually reviewing customer feedback or analysis of survey responses. However, such an approach is not feasible for OCR due to the large volume of text generated. Therefore, we use bi-gram analysis, a frequency distribution of co-occurring words in a string, on the subset of positive/negative review sentences associated with a topic to infer the common reasons for passenger satisfaction/dissatisfaction. Such an approach has been used in prior research to obtain deeper insights from OCR in industries such as logistics \cite{Rajendran2020} and education \cite{Srinivas2019}.

\section{Research Methodology}
\begin{figure}[h]
	\centering
	
	\captionsetup{justification=centering}
	\includegraphics[width=1\linewidth]{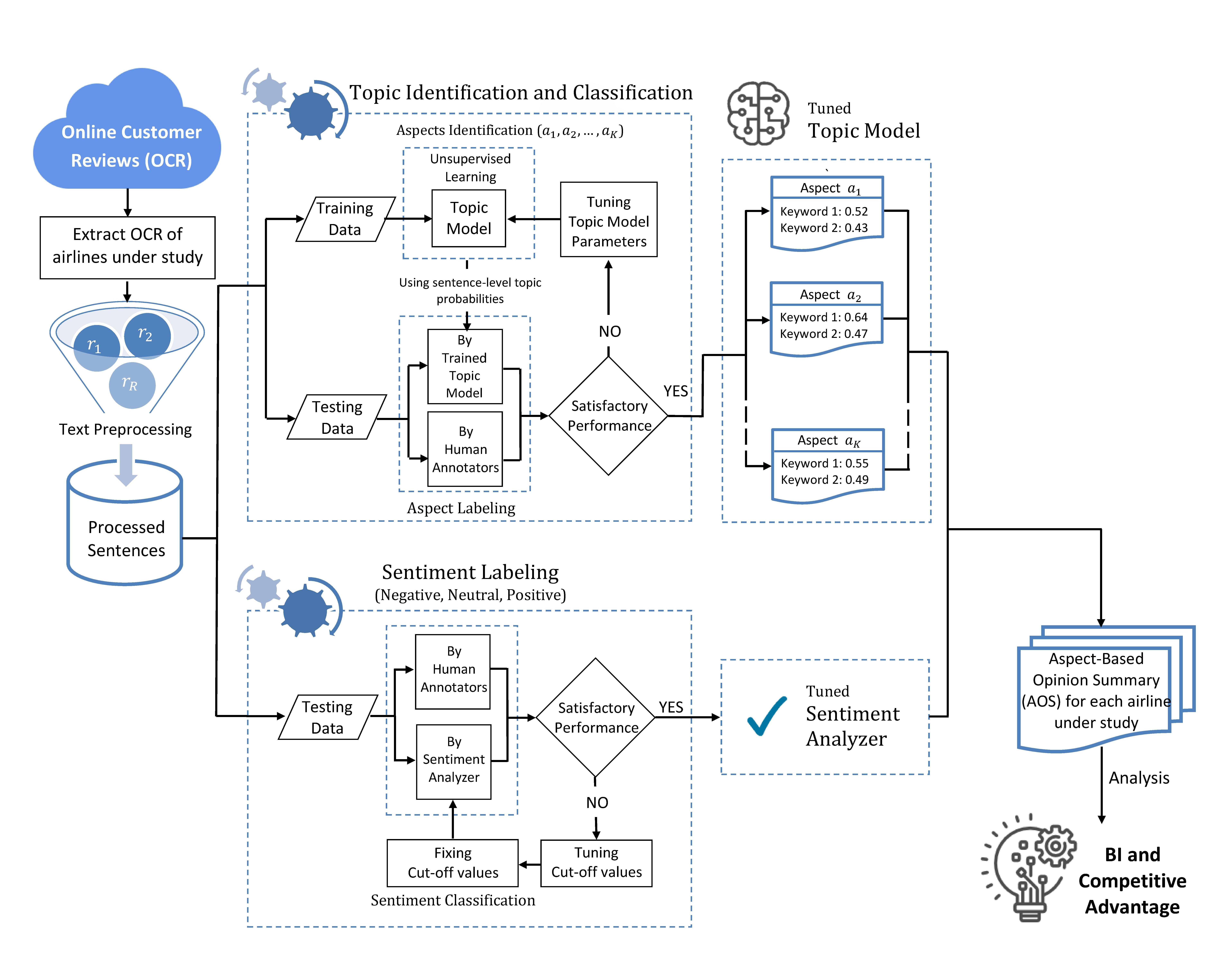}
	\caption{Overview of research methodology}
	\label{fig:Fig2}
\end{figure}

In this research, we strive  to derive business intelligence for an airline (referred to as the target airline), which seeks to understand passengers’ perspectives and improve their satisfaction, by analyzing OCR on itself and its competitors. To achieve this, we propose the research framework illustrated in Figure \ref{fig:Fig2}. First, the unstructured OCR of the target airline and its competitors are extracted using a web crawler and scraper. For each airline under consideration, the corpus containing all the customer reviews is split into individual sentences and then pre-processed to facilitate unsupervised learning. A sizeable  subset of the pre-processed text is provided as an input to  the topic models (such as pLSA and LDA) for identifying the prominent aspects discussed by the customers and to prepare the models for topic-based sentence stratification. On the contrary, the remaining text is held-out for extrinsic evaluation of the model’s capability in sentence classification. Upon completion of the un-supervised learning, the dominant topics in the OCR are established and the  probability of each topic to represent the individual  review sentences are  obtained, and this information is used for the discriminative task of labeling each sentence in the testing dataset with  its most probable aspect. The topic model’s performance is evaluated by comparing the automatically labeled sentences with the ground-truth (i.e., manually annotated sentences). Subsequently, the sentences are also labeled using sentiment analysis (SA) tools, and the performance of the approach adopted is evaluated. While the performance of a topic model is  optimized by tuning its hyperparameters, such as the number of topics to extract , SA tools typically do not need it.  SA tools, such as  SentiStrength \cite{Thelwall2017}, VADER \cite{Hutto2014}, and AFFIN \cite{Nielsen2011} are pre-trained on a large domain-independent corpus and are therefore directly employed on the held-out review sentences to obtain customer sentiments.  The SA tools employed provide a sentiment polarity score on a numerical scale, and a cut-off/threshold is determined  to convert it to a sentiment class (positive, negative or neutral) as well as to fine tune the model’s classification ability. The performance of a sentiment analyzer is evaluated by comparing its results with the ground-truth sentiments. The tuned topic model and sentiment analyzer are then leveraged to classify the review sentence of each airline into a topic and sentiment, respectively. Thereafter, the sentiments across all the review sentences discussing the same aspect of an airline are combined to obtain aspect-based opinion summary (AOS) of that carrier. Finally, business intelligence is obtained by analyzing the AOS of all carriers. Further, bi-gram based root-cause is executed on the categorized reviews to extract more specific-actionable insights. A detailed description of key stages associated with the proposed framework is provided in the following subsections.

\subsection{Passenger Review Extraction and Pre-Processing}
\begin{figure}[h]
	\centering
	
	\captionsetup{justification=centering}
	\includegraphics[width=1\linewidth]{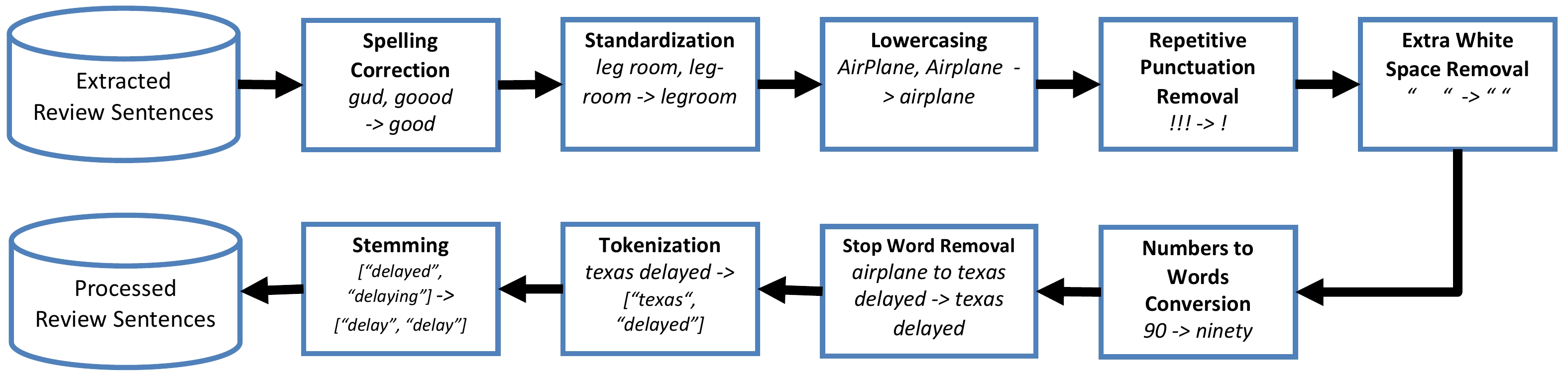}
	\caption{Overview of research methodology}
	\label{fig:Fig2a}
\end{figure}

The dataset can be retrieved from one or more online travel platforms (e.g., TripAdvisor, Kayak) that allow its registered users to post detailed textual reviews describing their experiences/opinions with an airline. As customer reviews are prone to noise and inconsistencies, the extracted OCR are pre-processed by employing different techniques, as shown in Figure \ref{fig:Fig2a}. Since a review can discuss multiple aspects and express different sentiment towards each aspect (as shown in Figure \ref{fig:Fig1}), it is first decomposed into individual sentences. Subsequently, a series of normalization procedures are  conducted to ascertain  consistency among all the reviews. This includes spelling correction, standardization of identical words, conversion to lowercase for uniformity, removal of repetitive punctuations and extra white spaces.  Following normalization, the stop words (e.g., “and”, “the”, “is”) are expunged  from the review sentences as  they occur very frequently while not adding any context/value to our models. Every review sentence is then broken down into a set of words (tokenization) and the words are reduced from their inflectional forms to  base forms (stemming) to enable better  semantic interpretations. Finally, the pre-processed review sentences corresponding to all the airlines are partitioned into learning and held-out data.

\subsection{Aspect Extraction using Topic Modeling}
The review sentences in the learning data are used as inputs to topic models, which automatically uncover the latent variables (referred to as topics or aspects) that govern the semantics of the corpus. The topic models assume the corpus containing $R$ review sentences, $C = \{1,2,…,R\}$, to be a mixture of a finite set of $K$ aspects, $A = \{a_1,a_2,…,a_K\}$, where each aspect $(a_k \in A)$ is characterized by a multinomial distribution of words \cite{Blei2003}. Further, there are $U$ unique words, $W =  \{w_1,w_2,w_3,…,w_U\}$, in corpus $C$ and each passenger review sentence, $r \in C$, contains a bag of words ($B$) with word sequence $\{w_1,w_2,w_3,…,w_B\}$, where $w_i$ denotes the $i^\text{th}$ word in the sequence. During training, a topic model uses a generative procedure to determine the aspect distribution in each review sentence $r$ (i.e., $P(a_k|r), a_k \in A;r \in R)$ and word distribution in each aspect (i.e., $P( w|a_k), w \in W,a_k\in A)$. Depending on the generative procedure and statistical assumptions employed in the  topic modeling algorithm, both $P( w|a_k)$  and $P(a_k|r)$ may change. Among the various topic modeling algorithms, pLSA \cite{Hofmann2001} and LDA \cite{Blei2003} are widely utilized  in the literature \cite{Lu2011}. Therefore, we use these algorithms for identifying the topics and the associated keywords. The generative procedure of pLSA, and LDA are briefly described in the following subsections.

\subsubsection{Probabilistic latent semantic analysis (pLSA)}
Probabilistic latent semantic analysis (pLSA) models each word in the review sentence as a sample from a mixture model with multinomial component models and is represented as shown in Equation (\ref{eq:1}) \cite{Hofmann2001}.

\vspace{-1cm}
\begin{align}
& P(w|r)= \sum_{a=1}^k P(w|a)P(a|r)   \label{eq:1}
\end{align}

The word-aspect distributions $P(w|a)$ and aspect-review distributions $P(a|r)$ are evaluated using the Expectation-Maximization (EM) algorithm \cite{Dempster1977} that maximizes the likelihood that the review corpus ($C$) is generated by the model and is given by Equation (\ref{eq:2}). Note that the parameters $\phi$ and $\theta$ indicate the words that are important for a certain aspect and aspects that appear in  a review sentence, respectively. Also, $c(w,r)$ denotes the number of times a word $w$ occurs in a review sentence $r$.

\vspace{-1cm}
\begin{align}
  & \text{log} \, P(r|\phi,\theta) =\sum_{r\in C} \sum_{w \in W} \{c(w,r) \, \text{log} \sum_{a \in A}  P(w|a)P(a|r)\}
 \label{eq:2}
\end{align}

The expectation step involves the evaluation of the hidden variable $t$ using the model parameters evaluated in the previous iteration and is given by Equation (\ref{eq:3}):

\vspace{-1cm}
\begin{align}
& P(t_{r,w} = a) =  \dfrac{P(w|a) P(a|r) }{\sum_{a^\prime \in A}P(w|a^\prime) P(a^\prime|r)}   \label{eq:3}
\end{align}

The maximization step involving updating formulas for model parameters are given by Equations \ref{eq:4} and \ref{eq:5}.

\vspace{-1cm}
\begin{align}
& P(a|r)=  \dfrac{\sum_{w \in W}  c(w,r)  P(t_{r,w} = a)}{\sum_{a^\prime} \sum_{w\in W}  c(w,r)  P(t_{R,w} = a^\prime)}  \label{eq:4}
\end{align}

\vspace{-1cm}
\begin{align}
& P(w|a)=  \dfrac{\sum_{r\in C}  c(w,r)  P(t_(r,w)=a)}{\sum_{w^\prime \in W}  \sum_{r\in C}  c(w^\prime,r)  P(t_{r,w^\prime}=a)} \label{eq:5}
\end{align}

\subsubsection{Latent Dirichlet Allocation (LDA)}
Though pLSA model is effective in extracting topics, it is prone to over fitting and does not provide an explicit way to process new review documents that aren’t a part of the training data. In order to overcome these challenges, Blei et al. \cite{Blei2003} proposed the Latent Dirichlet Allocation (LDA), a topic model that closely resembles PLSA but majorly differs in the way it makes statistical inferences. Specifically, LDA derives its topic mixture ($\theta$) from a conjugate Dirichlet prior. The generative process for each review sentence $r$ in the corpus $C$ is stated below, where $\alpha$ and $\beta$ are hyper-parameters of the Dirichlet and are determined beforehand based on the review documents’ aspect distribution and word distribution, respectively; $\theta$ is the data generating $N$-dimensional Dirichlet random variable.
\begin{enumerate}
	\item Choose $U \sim Poisson(\epsilon)$
	\item Choose $\theta \sim Dir(\alpha)
$
	\item For each of the $B$ words $w_i$:
	\begin{itemize}
		\item Sample an aspect $a_i \sim Multinomial(\theta)$
	\item Sample word $w_i$ from $P(w_i|a_i,\beta)$, a multinomial probability based on the aspect $a_i$
	\end{itemize}
\end{enumerate}

The aspect distribution ($\theta$) conditioned on the hyperparameter $\alpha$ is given by Equation \ref{eq:6}.

\vspace{-1cm}
\begin{align}
& P(\theta|\alpha)=  \dfrac{\Gamma (\sum_{k=1}^K \alpha_k)}{\Pi_{k=1}^K \Gamma(\alpha_k)} \theta_1^{\alpha_1-1},….,\theta_K^{\alpha_k-1}
 \label{eq:6}
\end{align}

Given the parameters $\alpha$ and $\beta$, the joint distribution of aspect mixture ($\theta$), a set of $B$ aspects $A$, and a set of $B$ words $r$ is given by Equation \ref{eq:7}.

\vspace{-1cm}
\begin{align}
& P(\theta,A,r|\alpha,\beta)=P(\theta|\alpha)  \Pi_{i=1}^B  P(a_i|\theta)  P(r_i|a_i,\beta)
\label{eq:7}
\end{align}

A review sentence’s marginal distribution obtained by integrating over $\theta$ is defined by Equation \ref{eq:8}.

\vspace{-1cm}
\begin{align}
& P(r|\alpha,\beta)= \int P(\theta|\alpha)  \big(\Pi_{i=1}^B \sum_{a_i}  P(a_i|\theta)  P(r_i|a_i,\beta)\big)d\theta \label{eq:8}
\end{align}

Finally, the corpus’s probability obtained by taking the marginal probabilities of all review documents is given by Equation \ref{eq:9}.

\vspace{-1cm}
\begin{align}
& P(C|\alpha,\beta)=\Pi_{r=1}^R  \int P(\theta_r |\alpha) (\Pi_{i=1}^{B_r} \sum_(a_{r,i})  P(a_{r,i}|\theta_r)  P(w_{r,i}|a_{r,i},\beta))d\theta_r 
\label{eq:9}
\end{align}

To adopt the LDA, it is necessary to compute the posterior distribution of the hidden variables given a review sentence. While it is intractable for exact inference, other approximate algorithms can be leveraged to solve the posterior distribution. In this research, we consider two approximate inference algorithms for LDA, namely variational approximation and Markov Chain Monte Carlo algorithms (specifically  Gibbs Sampling). Thus, two variants of LDA are considered:  LDA based on variational inference (LDA-VI) and LDA based on Gibbs Sampling (LDA-GS)

\subsection{Determining Number of Aspects and Aspect Labeling}

The inference from the topic model includes a sequence of keywords arranged in the decreasing order of their likelihood of representing a particular aspect. One of the challenging tasks is to designate  the number of topics ($K$) to  establish without having prior knowledge on the aspects of service quality discussed in the OCR \cite{Pavlinek2017}. The topic model becomes too coarse to detect the key dimensions of service quality discussed by the passengers if $K$ is underestimated. On the other hand, overestimating the aspects could lead to complex topic models, thereby making it arduous to validate and implement. A prevalent  statistical technique to select $K$ is to estimate the predictive likelihood of the topic model on held-out data using the perplexity measure \cite{Chang2009}. However, such an approach may not always reveal human interpretable topics, which are necessary for applications such as topic segmentation \cite{Chang2009}.  Therefore, we use the topic coherence score, an evaluation metric that measures the degree of semantic similarity between high probability keywords in the topic, to determine the optimal number of topics.  The Coherence for topic $k$ containing $n$ probable words $(w_1, w_2,…,w_n)$ is given by Equation \ref{eq:10}, where $P(w_i)$ denotes the likelihood of finding $w_i$ in a random document and $P(w_i,w_j)$ represent the probability of locating both $w_i$ and $w_j$ in an arbitrary   document.

\vspace{-1cm}
\begin{align}
& Coherence_k = \dfrac{\sum_{i=1}^{n-1}\sum_{j=i+1}^{n} \text{log} \dfrac{P(w_i,w_j)}{P(w_i) P(w_j)} }{{n \choose 2}}\label{eq:10}
\end{align}

For each topic model, the number of topics is varied from $K^{\text{min}}$ to $K^{\text{max}}$, and the corresponding coherence score is calculated. The optimal number of aspects is the smallest value of $K$ that yields the highest coherence score. Since topic models do not automatically label these aspects, we use our domain knowledge to label them by examining the relationship between the keywords associated with each aspect.  

\subsection{Review Sentence Classification}
The sentence-level topic distribution obtained from pLSA, LDA-VI and LDA-GS are used for the discriminative task of tagging the review sentences to their  most representative aspects. The procedure for classifying a review sentence $r$ using a trained topic model is illustrated in Equation \ref{eq:11}. 

\begin{equation}\label{eq:11}
T_r = 
\begin{cases}
\underset{k}{\text{argmax}} \, \{P(a_k|r) : k = 1,2,...,K\} & \text{if } \underset{k}{\text{max}}\, \{P(a_k|r) : k = 1,2,...,K \} > \gamma \\
\text{Null},              & \text{otherwise}
\end{cases}
\end{equation}

%\underset{k}{\text{argmax}}⁡\, 
%\underset{k}{\text{max}}⁡ 

A review sentence $r$ is assigned to the most probable topic only if its corresponding probability is greater than a user-specified threshold $\gamma$. Otherwise, it is not assigned to any of the $K$ topics and is instead categorized as ‘Null’ topic. A threshold is used to avoid classification of objective statements which do not reflect customer’s perception of service quality. For example, sentences such as “\textit{I traveled from Los Angeles to San Francisco}” or “I\textit{ chose to book with this airline in March}” are objective statements and should be classified as ‘Null’ topic for the purpose of this research. Besides, the aspects identified by the topic models contain  only the  commonly discussed themes  and thus are not exhaustive. Therefore, it is possible to encounter review sentences that do not belong to the identified $K$ topics and its crucial to categorize them appropriately.  The provision for classifying a review sentence as ‘Null’ also helps to overcome this challenge. Note that a high threshold value will increase the accuracy of tagged sentences but may discard  many sentences by classifying them as ‘Null’. On the other hand, a low $\gamma$ will lead to increased misclassifications.  Further, the accuracy is also challenged due to the presence of non-coherent keywords associated with an aspect or keywords reflecting multiple topics. 
	
To improve the sentence classification procedure, we propose an ensemble assisted topic model (EA-TM), a meta-algorithm that combines $M$ topic models, $m = \{1,2,…,M\}$, such that the accuracy achieved is better than the individual topic models. Specifically, a rule-based approach is adopted to predict the most appropriate topic for a given review sentence. First, the EA-TM seeks to classify a sentence by plurality voting of the most probable aspects identified by individual topic models, provided there exists a topic that is unanimously chosen by the majority of the topic models under consideration (i.e., a unique mode). If $\tau_{r,m}$ represents the most probable topic chosen by model $m$ for sentence $r$ (i.e., ($\tau_{r,m} = \underset{k}{\text{argmax}} \, {P(a_k|r,m) : k=1,2,…,K}$) and $\{\tau_{r,m}:m=1,2,…,M\}$ is unimodal, then the classification of sentence $r$ by EA-TM can be represented as shown in Equation \ref{eq:12}. 

\vspace{-1cm}
\begin{align}
& T_r=mode\,\{\tau_{r,m}:m=1,2,…,M\}
\label{eq:12}
\end{align}

However, if there is a tie in voting (multimodal) or when there is no mode, the review sentence ($r$) is allotted the aspect labeled by the most confident of the individual topic models, $T_r=\underset{k}{\text{argmax}} \, \{P(a_k|r,m) : k=1,2,…,K;m=1,2,…,M\}$, provided the associated topic probability is greater than $\gamma$. 

Nevertheless, both  plurality voting and threshold-based approaches may at times fail to automatically assign  a topic, even when the  sentence belongs to one of the established $K$ topics. This could happen when words mentioned in the review sentence do not occur frequently in the corpus. For example, the sentence “The burger was stale” clearly reflects a topic associated with “Food/In-flight Meals”, but the keywords generated by the topic model may not include the word ‘burger’ as it could have appeared only in a few review sentences. To facilitate the classification of such statements, a topic-specific custom word list, $\{L_k: k=1,2,…,K\}$, is generated by automatically extracting all the synonyms, antonyms, hyponyms, and hypernyms associated with each keyword identified by the topic model. Subsequently, a score for classifying a review sentence $r$ to an aspect $a_k$ , $S(a_k)$, is computed by counting the number of words from $r$ that appear in the custom word list $L_k$. A sentence, not satisfying the proposed condition-based approaches, is now classified into the  topic with the highest score as shown in Equation \ref{eq:13}, while ties in the scores are resolved by arbitrarily choosing one of the highest scoring aspects.

\vspace{-1cm}
\begin{align}
& T_r=\underset{k}{\text{argmax}} \, {S(a_k):k=1,2,…,K})
\label{eq:13}
\end{align}

If the aforementioned rules are unable to classify a sentence to a most appropriate topic, then the sentence  is labeled as a ‘Null’ topic. In summary, the classification of review sentence $r$ by EA-TM is discussed in Algorithm \ref{Alg:EA-TA}.

\begin{algorithm}[h]
	\caption{Sentence-level Aspect Classification using Ensemble-assisted Topic Model}
	\label{Alg:EA-TA}
	\begin{algorithmic}[1]
		\State \textbf{Input:} Review sentences $\{r : r = 1,2,...,R\}$,  identified topics $\{a_k:k=1,2,...,K\}$, topic-specific custom word-lists $\{L_k: k=1,2,…,K\}$ and threshold parameter ($\gamma$)
		\State \textbf{{Output:}} Topic associated with each review sentence $r$ ($T_r$)
		\State \textbf{Procedure:}
		\State \textbf{Initialize} \{$T_r = Null:r = 1,2,...,R$\}
		\For{each review sentence $r \in \{1,2,...,R\}$}
		\State \textbf{Set} \{$S(a_k) = 0: k = 1,2,...,K$\} 
				\For{each topic model $m \in \{1,2,...,M\}$}
				%\State {Identify the most probable aspect for sentence $r$ using topic model $m$}
				\State $\tau_{r,m} = \underset{k}{\text{argmax}} \,\{P(a_k|r,m): k = 1,2,...,K\}$ 
					\EndFor		
					
					\If{$\{\tau_{r,m} : m=1,2,…,M\}$ is unimodal}
					\State $T_r = \text{mode}\{\tau_{r,m} : m=1,2,…,M\}$
					\ElsIf{$\underset{k}{\text{max}} \,\{P(a_k|r,m): k = 1,2,...,K; m=1,2,...,M\} > \gamma$}
					\State  $T_r = \underset{k}{\text{argmax}} \,\{P(a_k|r,m): k = 1,2,...,K; m=1,2,...,M\}$
			\Else
			\For{each topic $k \in \{1,2,...,K\}$}
			\For{each word $w$ in sentence $r$}
					\If{$w$ is in list $L_k$}
						\State $S(a_k) = S(a_k) +1$
					\EndIf
			\EndFor
			\EndFor
			\State  $T_r = \underset{k}{\text{argmax}} \, \{S(a_k):k=1,2,…,K\}$
					\EndIf
					
						\EndFor		
	
	\end{algorithmic}
\end{algorithm}

\subsection{Sentiment Analysis of Review Sentence} 
Besides tagging a sentence with a topic, identifying the customer perception associated with each of these review sentences is crucial to obtain an aspect-level sentiment summary. For this purpose, we use sentence level-sentiment analysis to  facilitate the categorization of customer perception as positive, neutral or negative. Owing to its wide and crucial applications, multitudes of researches have taken place to introduce numerous sentiment analysis techniques. Ribeiro et al. \cite{Ribeiro2016} studied the performance of 24 popular sentiment analysis methods across gold standard labeled data sets covering different domains. They found three lexicon-based methods, namely, SentiStrength \cite{Thelwall2017}, Valence Aware Dictionary for sEntiment Reasoning or VADER \cite{Hutto2014} and AFFIN \cite{Nielsen2011} to consistently rank well for datasets involving customer reviews. The overall sentiment score for a review sentence is typically obtained by combining  the individual sentiment scores for each word in that sentence.  AFFIN sentiment evaluator has 2477 coded words in its dictionary, where each word has a valence score ranging from -5 (unpleasant) to 5 (pleasant). It is designed for modern English and therefore can efficiently extract sentiment from review sentences involving internet slang, web jargons and obscene words. On the other hand, SentiStrength is a lexicon of 2310 words and provides a scaled sentiment score (varying from -1 to +1) for review sentences. In particular, it is effective for handling sentences containing exaggerated punctuations, emoticons, and deliberate misspellings. Finally, VADER, which includes 7,500 lexical features (words, slang, emojis, etc.) that are specially adapted for social media data, rates the sentiment on a -1 (extremely negative) to +1 (extremely positive) continuous scale. Besides using a comprehensive lexicon, VADER uses rules to capture the grammatical and syntactical conventions used to express sentiment intensity. Yet, the performance of these three opinion mining methods varied substantially across different sentences owing to differences in their lexical features and rule-based scoring \cite{Ribeiro2016}. Therefore, to obtain consistent sentiment orientation of the review sentences and exploit the strengths of the three well-performing SA methods, this research introduces an ensemble sentiment analysis (E-SA) algorithm that combines the predictions from these methods to produce superior results spanning across diverse customer reviews. The step-by-step procedure for E-SA is presented in Algorithm \ref{Alg:E-SA}. Once the sentiment score for sentence $r$ is determined by sentiment evaluator l ($\nu_{r,l}$), the E-SA approach leverages positive and negative sentiment cut-off values ($\gamma^+$  and $\gamma^-$) to convert $\nu_{r,l}$ into a “positive”, “negative” or “neutral” sentiment category ($o_{r,l}$). Subsequently, a plurality voting of $L$ individual sentiment analyzers is used to determine the sentiment associated with each review sentence $r$ ($O_r$) as shown in Equation \ref{eq:14}.

\vspace{-1cm}
\begin{align}
& O_r= mode\, \{o_{r,l}:r=1,2,…,R; l=1,2…,L\}
\label{eq:14}
\end{align}

In case of tie or absence of a mode, E-SA uses a three-step procedure to ascertain the sentiment category: (i) normalizes the scores obtained for sentence $r$ by each $l$ between -1 and 1 ($\nu^\prime_{r,l}$) (ii) selects the sentiment analyzer that has the highest absolute value of the normalized sentiment score (denoted as $l^*$), (iii) establishes the sentiment of sentence $r$  to reflect the sentiment category predicted  by the most confident model (i.e., $O_r = o_{r,l^*}$).

\begin{algorithm}[h]
	\caption{Sentence-level Sentiment Classification using Ensemble Sentiment Analyzer}
	\label{Alg:E-SA}
\begin{algorithmic}[1]
		\State \textbf{Input:} Review sentences $\{r: r = 1,2,...,R\}$ and sentiment threshold values  ($\gamma^+$,    $\gamma^-$) 
		\State \textbf{{Output:}} Sentiment category associated with each review sentence $r$ ($O_r$)
		\State \textbf{Procedure:}
		\State \textbf{Initialize} \{$O_r = Null:r = 1,2,...,R$\}
		\For{each review sentence $r \in \{1,2,...,R\}$}
	
		\For{each sentiment analysis method $l \in \{1,2,...,L\}$}
				\State Obtain sentiment score ($v_{r,l}$)
				\State Convert $v_{r,l}$ to sentiment category $o_{r,l}$
				\If{$v_{r,l} > \gamma^+_l$}
				\State $o_{r,l} = \text{Positive}$
				\ElsIf{$\gamma^-_l\leq v_{r,l} \leq \gamma^+_l$} 
				\State $o_{r,l} = \text{Neutral}$
				\Else
					\State $o_{r,l} = \text{Negative}$
				\EndIf
		\EndFor

			\If{$\{o_{r,l} : l=1,2,…,L\}$ is unimodal}
			\State $O_r= \text{mode}\{o_{r,l}: l=1,2…,L\}$
			\Else
			\For{each sentiment analysis method $l \in \{1,2,...,L\}$}
%
%max(v_{r,l^\prime}:l^\prime=1,2,...,L) - min(v_{r,l^\prime}:l^\prime=1,2,...,L)
\State $v_{r,l}^\prime  = 2\frac{ v_{r,l} - \min(v_{r,l^\prime}: r^\prime=1,2,..,R)}{\max(v_{r,l^\prime}:r^\prime=1,2,...,R) - \min(v_{r,l^\prime}:r^\prime=1,2,...,R)} - 1$

			\EndFor		
		\State $l^* = \argmax_l\{|v_{r,l}^\prime|: l=1,2…,L\}$	
		 \State $O_r= o_{r,l^*}$
			
		\EndIf
		\EndFor		
	
	\end{algorithmic}
\end{algorithm}

\subsection{Evaluation of Aspect-detection and Sentiment Classification}
To evaluate and compare the performance of the topic-models and sentiment analyzers extrinsically, we first compare the results inferred by each of the models with respect to the ground truth. The  benchmark models as well as the proposed ensemble models are each used to label the review sentences in the testing data set, which contains the held-out reviews that have not been used for training. Specifically, pLSA, LDA-VI, LDA-GS, and EA-LDA are separately  used to label the aspects for the held-out review sentences, while AFINN, SentiStrength, VADER and E-SA are individually  used to classify the associated sentiment category. The ground truth is established with the help of human annotators, who are presented with testing dataset as well as the identified aspects and sentiment categories. In order to avoid bias, multiple human annotators are employed to tag the topic and sentiment associated with each sentence. In case of discrepancies between annotators, that particular review sentence is analyzed again to resolve the discrepancy in classification. In order to quantitatively evaluate and compare the performances, we calculate three commonly used measures in the literature - aspect-wise precision, recall and $F_1$ scores. Precision ($\text{P}$) is the ratio of the number of correct classifications to the total number of instances where the model inferred it. It is preferred  when the cost of incorrectly classifying a review into that aspect or sentiment is high. On the other hand, recall ($\text{R}$) is the ratio of the number of correct classifications to total number of its instances in the ground truth. It is adopted when the cost of failing to classify an appropriate review into that aspect or sentiment is high. Finally, the $\text{F}_1$ score is the harmonic mean of precision and recall, as shown in Equation \ref{eq:15}.

\vspace{-1cm}
\begin{align}
& \text{F}_1= 2\dfrac{\text{P} \times \text{R}}{\text{P} + \text{R}} \label{eq:15}
\end{align}

$\text{F}_1$ score is better metric to consider when we seek a balance between Precision and Recall. The maximum value for all three evaluation metrics is 1, and is obtained when a model has classified all the instances correctly.

\subsection{Aspect-level Sentiment Analysis for Business and Competitive Intelligence}
For each of the airlines under consideration, the AOS is  obtained by combining  review sentences  on the same aspect and stratifying them based on their sentiment categories. Given the AOS of the target airline, a topic that has a stupendously  positive overall sentiment is  regarded as a customer-perceived strength, while an aspect with an overall negative sentiment is its weakness. Besides, to obtain the key  reasons for customer appreciations and criticisms pertaining to the target airline, a bi-gram analysis is conducted on the aspects considered as strengths and weaknesses, respectively. This enables the target airline to develop/plan specific actionable insights. Further, by evaluating the AOS of the competitors, insights on their strengths and weaknesses are discovered.  Such analysis helps an organization better understand their customer’s perceptions and plan managerial actions accordingly. Besides, it also aids in acquiring business intelligence – information needed to make profitable business decisions (i.e. intelligence on strengths and weaknesses of the target airline and its competitors). 

\section{Case Study}
In this section, we evaluate the feasibility of the proposed methodology using a case study. First, the parameters related to the case study are  established and the dominant topics discussed in the reviews are extracted. Subsequently, each review sentence is classified based on the relevant topic and sentiment, and the performance of the proposed EA-TM and E-SA models is  validated. Finally, the results are analyzed to gain business intelligence for the target airline, and its theoretical, design, and managerial implications are discussed. The research framework (e.g., data extraction, pre-processing, model development) was coded using Python programming language, which provides numerous libraries for text mining and NLP, and executed on a computer running Windows 10 with an i9 processor and 128 GB RAM. 

\subsection{Case Study Parameters}
We considered a full-service carrier (FSC), a major US-based airline operating a large domestic network spanning 242 destinations with over 160 million annual passengers and 5000 daily departures yet ranked only in lower the quartile for service quality when compared with other domestic carriers in the US, as our target airline. Additionally, we  considered four of its competitor airlines, including two LCC carriers (LCC-1 and LCC-2) and two FSC airlines (FSC-1 and FSC-2). The airlines were selected based on their service similarities and core competencies. While FSC airlines have traditionally dominated the US market, LCC airlines have grown tremendously in the last decade, and studying them could give insights into the service gaps that they might have capitalized.  Specifically, LCC-2 had achieved the highest improvement in service quality in recent years, while FSC-1, FSC-2 and LCC-1 dominated the domestic market share. Besides, FSC-2 and LCC-1 also ranked in the top quartile for service quality in the US. 

The dataset utilized in this paper was retrieved from TripAdvisor – the world’s largest travel platform that facilitates booking and provides user-generated feedback \cite{Korfiatis2019}. TripAdvisor has over 830 million reviews of 8.6 million hospitality services \cite{Tripadvisor2019}. Therefore, we capitalized on this freely available information and extracted all the reviews that were published between August 2017 and September 2019 for the five carriers. The web scraping was performed by employing the Beautiful Soup library in Python, and the Selenium WebDriver  was used to navigate through the web pages.  A total of 99,147 reviews, which had over 398,571 unique sentences with an average length of 18 words, were scraped. Subsequently, the review sentences were pre-processed using the functions available in Python’s Natural Language ToolKit (nltk) library. A majority of the extracted review sentences ($\sim$ 95\%) were utilized for unsupervised learning of the topics and sentiments.  Supervised learning was avoided as it requires manual annotation of the review sentences, which is time-consuming, cumbersome and expensive. However, the remaining 20,000 review sentences were held-out, but only to extrinsically evaluate the discriminative capability of the topic models and sentiment analyzers. The topic models, namely, pLSA, LDA-VI and LDA-GS were implemented using the libraries available in python. To establish the number of topics ($K$) in each topic model, we varied its value from 5 to 50 in increments of 1 and calculated the corresponding topic coherence score.

Upon extracting and labeling the topics discussed in the scrapped reviews using the three topic models, the individual topic models and E-TA were employed for review sentence classification. We observed the accuracy of sentence-level topic classification to be best with a threshold parameter ($\gamma$) of 0.7. Similarly, the individual sentiment analyzers and the E-SA were used for sentence-level sentiment classification and based on empirical evaluation, the positive ($\gamma^+$) and negative ($\gamma_-$) threshold values for both SentiStrength and VADER were  established as 0.05 and -0.05, respectively. On the other hand, the values of $\gamma^+$ and $\gamma^-$ were set to -1 and +1 for AFINN. %We then compared the performance of the proposed EA-TM with the individual topic models (i.e., pLSA, LDA-VI and LDA-GS) and the performance of the proposed E-SA with the individual benchmark SA methods (i.e., SentiStrength, VADER, and AFFIN) using Recall, Precision and $\text{F}_1$  scores.

\subsection{Experimental Results}
The best coherence value was achieved with  18 topics for pLSA and 23 topics for the other two topic models (LDA-VI and LDA-GS). Subsequently, the topics were labeled based on the keywords identified. Similar to Lucini et al. \cite{Lucini2020}, we grouped the topics that had similar keywords/meaning and discarded infrequent ones  (i.e., topic distribution probability of $<$ 0.05\%). Finally, we were able to identify 11 meaningful aspects from each of the three topic models. Table \ref{tab:Tab1} presents  the identified topics, along with their most representative words specific to each topic model. In addition to the expected aspects of service quality, such as seating and baggage services, the model also identified ticketing services and reward programs as frequently discussed topics in the passenger reviews. Moreover, the priority for each  keyword varied depending on the topic model. For instance, the highest ranked keyword pertaining to the “Ground and Cabin Staff” topic is “crew”, “staff” and “attendant” for pLSA, LDA-VI and LDA-GS, respectively. 

% Table generated by Excel2LaTeX from sheet 'Sheet1'
\begin{table}[htbp]
  \centering
  \caption{Topics inferred along with most representative keywords}
   \scalebox{0.6}{  \begin{tabular}{llll}
    \toprule
    \multicolumn{1}{c}{\textbf{Topic}} & \multicolumn{1}{c}{\textbf{LDA-VI Keywords}} & \multicolumn{1}{c}{\textbf{pLSA Keywords}} & \multicolumn{1}{c}{\textbf{LDA-GS Keywords}} \\
    \midrule
    Ground and Cabin Staff & staff, attendant, crew, service, customer  & crew, attendant, steward, employees, staff & attendant, professional, staff, friendly, helpful \\
    Inflight Entertainment & entertainment, movie, screen, app, wifi & movie, wifi, screen, device, entertainment & entertainment, free, movie, watch, system \\
    Rewards Program & miles, card, credit, frequent, flyer & miles, points, benefits, rewards, status  & rewards, frequent, mileage, points, silver \\
    Food and Beverage & food, snack, drink, meal, service & food, drink, snack, eat, meal & food, meal, drink, offered, water \\
    Check-in and Boarding & gate, boarding, line, board, pass & screening, boarding, security, gate, check & check, boarding, line, wait, group\\
    Seating & seat, space, cramped, seating, tight & seat, legroom, sitting, cramped, row & seat, row, aisle, window, sit \\
    First and Business Class & class, first, business, service, lounge & business, class, lounge, service, comfortable   & class, business, service, mile, upgrade \\
    Baggage Services & bag, luggage, check, carry, baggage & luggage, bag, carry, carousel, weight & baggage, checked, charge, luggage, pay \\
    Cabin Comfort & cabin, overhead, bathroom, storage, bin & bathroom, lavatory, smell, overhead, clean & cabin, clean, hot, blankets, cold \\
    On-time Performance & hour, delayed, time, late, connecting & late, hour, early, delay, technical & hour, delay, weather, connection, hotel \\
    Ticketing Services & ticket, called, phone, cancel, book  & book, voucher, overbook, fee, ticket & ticket, change, pay, booked, money \\
    \bottomrule
    \end{tabular}}%
  \label{tab:Tab1}%
\end{table}%%

To establish the ground-truth of topics in the held-out dataset, two annotators manually labeled each  review sentence with one of the 11 identified topics or a \textit{“Null”} topic. Furthermore, the sentiment associated with each review sentence was labeled positive, negative, or neutral. In order to assess the classification performance of the topic models, they were employed  to automatically classify each  held-out review sentence to a topic (based on Equation \ref{eq:11} for individual topic models and Algorithm 1 for EA-TM) and the evaluation measures, namely, precision, recall, and $\text{F}_1$  scores, for each topic, obtained by validating  the predicted outcome with the ground-truth, were compared. Table \ref{tab:Tab2} summarizes the performance of the considered benchmark models in  review sentence classification. It is evident that the proposed EA-TM approach achieved superior discriminative power for all the topics compared  to the individual topic models, especially with respect to $\text{F}_1$ score. Similarly, these performance measures for sentence-level sentiment classification were determined to evaluate the capability of the SA methods, and are  shown in Table \ref{tab:Tab3}. Similar to EA-TM, the proposed E-SA method consistently outperformed the individual SA methods. Thus, the results clearly demonstrate the dominance of the ensemble-based approaches for unsupervised classification of review sentence into the most prominent topic and sentiment.

% Table generated by Excel2LaTeX from sheet 'Sheet3'
\begin{table}[htbp]
	\centering
	\caption{Aspect classification performance of pLSA, LDA-VI, LDA-GS, and EA-TM on held-out data}
	\scalebox{0.8} {\begin{tabular}{l|ccc|ccc|ccc|ccc}
			\toprule
			\multicolumn{1}{c|}{\multirow{2}[4]{*}{\textbf{Topic}}} & \multicolumn{3}{c|}{\textbf{pLSA}} & \multicolumn{3}{c|}{\textbf{LDA-VI}} & \multicolumn{3}{c|}{\textbf{LDA-GS}} & \multicolumn{3}{c}{\textbf{EA-TM}} \\
			\cmidrule{2-13}          & \textbf{R} & \textbf{P} & \textbf{$\text{F}_1$} & \textbf{R} & \textbf{P} & \textbf{F1} & \textbf{R} & \textbf{P} & \textbf{$\text{F}_1$} & \textbf{R} & \textbf{P} & \textbf{$\text{F}_1$} \\
			\midrule
			Baggage Services & 0.51  & 0.83  & 0.63  & 0.63  & 0.84  & 0.72  & 0.48  & 0.74  & 0.58  & 0.71  & 0.95  & 0.81 \\
			Check-in and Boarding Process & 0.55  & 0.75  & 0.64  & 0.77  & 0.65  & 0.71  & 0.61  & 0.71  & 0.65  & 0.93  & 0.81  & 0.87 \\
			Cabin and Ground Staff & 0.89  & 0.71  & 0.79  & 0.78  & 0.7   & 0.74  & 0.78  & 0.58  & 0.67  & 0.9   & 0.83  & 0.86 \\
			Cabin Comfort & 0.53  & 0.79  & 0.63  & 0.52  & 0.87  & 0.65  & 0.72  & 0.9   & 0.8   & 0.72  & 0.91  & 0.8 \\
			First and Business Class Service & 0.6   & 0.77  & 0.67  & 0.66  & 0.74  & 0.7   & 0.68  & 0.8   & 0.73  & 0.89  & 0.74  & 0.81 \\
			Food and Beverage Service & 0.55  & 0.78  & 0.64  & 0.65  & 0.78  & 0.71  & 0.72  & 0.8   & 0.76  & 0.77  & 0.92  & 0.84 \\
			In-flight Entertainment & 0.74  & 0.65  & 0.69  & 0.79  & 0.64  & 0.7   & 0.68  & 0.6   & 0.64  & 0.74  & 0.87  & 0.8 \\
			On-time Performance & 0.71  & 0.76  & 0.73  & 0.73  & 0.76  & 0.74  & 0.66  & 0.7   & 0.68  & 0.89  & 0.83  & 0.86 \\
			Rewards Program & 0.41  & 0.79  & 0.53  & 0.56  & 0.7   & 0.62  & 0.7   & 0.79  & 0.74  & 0.72  & 0.93  & 0.81 \\
			Seating & 0.87  & 0.71  & 0.78  & 0.72  & 0.81  & 0.77  & 0.56  & 0.74  & 0.63  & 0.79  & 0.91  & 0.85 \\
			Ticketing Services & 0.73  & 0.8   & 0.76  & 0.78  & 0.65  & 0.71  & 0.71  & 0.51  & 0.59  & 0.91  & 0.73  & 0.81 \\
			NULL  & 0.68  & 0.33  & 0.44  & 0.85  & 0.5   & 0.63  & 0.73  & 0.49  & 0.59  & 0.85  & 0.57  & 0.68 \\
			\bottomrule
	\end{tabular}}%
	\label{tab:Tab2}%
\end{table}%

% Table generated by Excel2LaTeX from sheet 'Sheet3'
\begin{table}[htbp]
  \centering
  \caption{Sentiment classification performance of SentiStrength, VADER, AFINN and E-SA on held-out data}
   \scalebox{0.8}{\begin{tabular}{lccc|ccc|ccc|ccc}
    \toprule
    \multicolumn{1}{c}{\multirow{2}[4]{*}{\textbf{Sentiment}}} & \multicolumn{3}{c|}{\textbf{SentiStrength}} & \multicolumn{3}{c|}{\textbf{VADER}} & \multicolumn{3}{c|}{\textbf{AFINN}} & \multicolumn{3}{c}{\textbf{E-SA}} \\
\cmidrule{2-13}          & \textbf{R} & \textbf{P} & \textbf{$\text{F}_1$} & \textbf{R} & \textbf{P} & \textbf{$\text{F}_1$} & \textbf{R} & \textbf{P} & \textbf{F} & \textbf{R} & \textbf{P} & \textbf{$\text{F}_1$} \\
    \midrule
    Positive & 0.67  & 0.87  & 0.76  & 0.75  & 0.85  & 0.8   & 0.86  & 0.76  & 0.81  & 0.88  & 0.93  & 0.9 \\
    Negative & 0.72  & 0.91  & 0.81  & 0.82  & 0.74  & 0.78  & 0.65  & 0.84  & 0.73  & 0.85  & 0.92  & 0.88 \\
    Neutral & 0.84  & 0.4   & 0.54  & 0.67  & 0.75  & 0.71  & 0.79  & 0.54  & 0.64  & 0.85  & 0.62  & 0.72 \\
    \bottomrule
    \end{tabular}}%
  \label{tab:Tab3}%
\end{table}%

\begin{figure}[h]
	\centering
	\captionsetup{justification=centering}
	\includegraphics[width=1\linewidth]{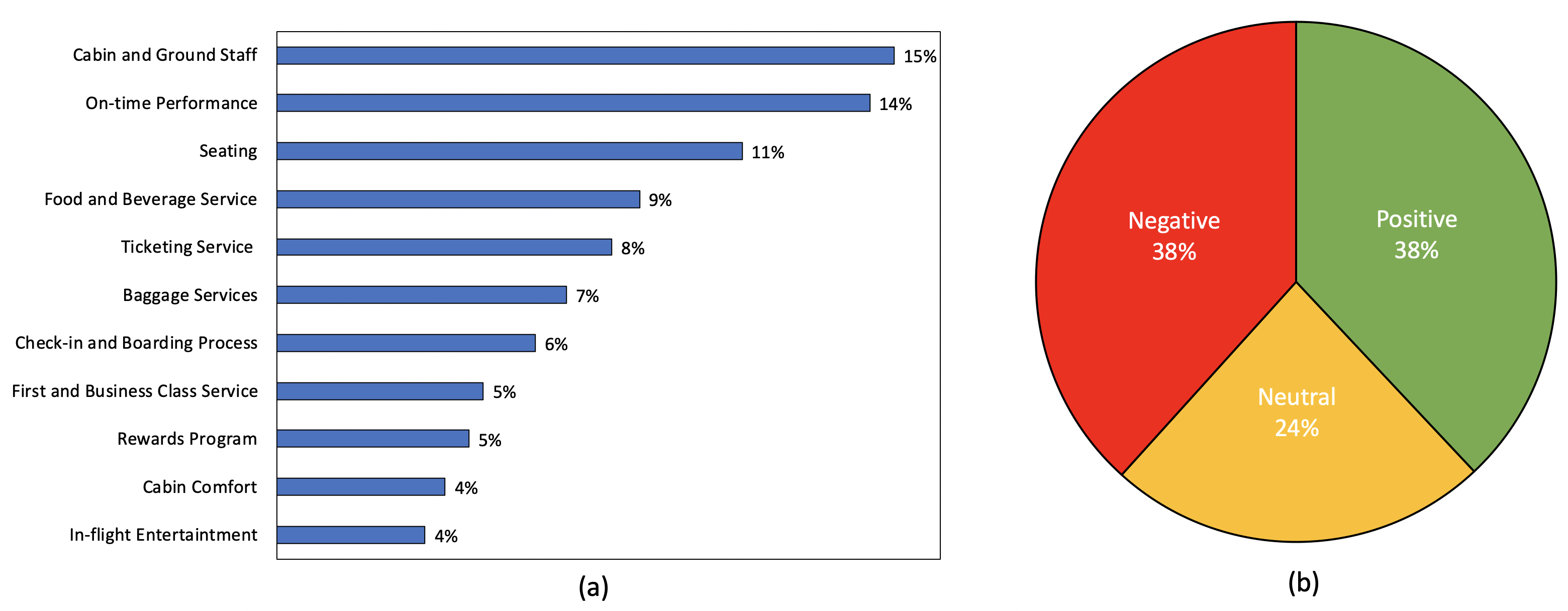}
	\caption{Overall distribution of (a) topics and (b) sentiment in the review sentences}
	\label{fig:Fig3}
\end{figure}

After validating the performance of the proposed models on the held-out data, we utilized  the trained EA-TM and E-SA to automatically classify the remaining 79,147  review sentences into  the most probable topic and sentiment, respectively. Figure \ref{fig:Fig3} provides the overall topic and sentiment distribution of the review sentences for the airlines under consideration. It can be observed that 40\% of the review sentences discuss specifically about staff, on-time performance or seating services provided by the airlines, thereby indicating it to be the critical dimensions of service quality for passengers. Besides, the distribution of positive and negative sentiments is balanced across the review sentences. Further, to enable business intelligence for the target airline, the sentence-level sentiments were categorized for each topic and carrier, and the airline-specific AOS were established.

\subsection{Extracting Business Intelligence}
\begin{figure}[h]
	\centering
	\captionsetup{justification=centering}
	\includegraphics[width=0.8\linewidth]{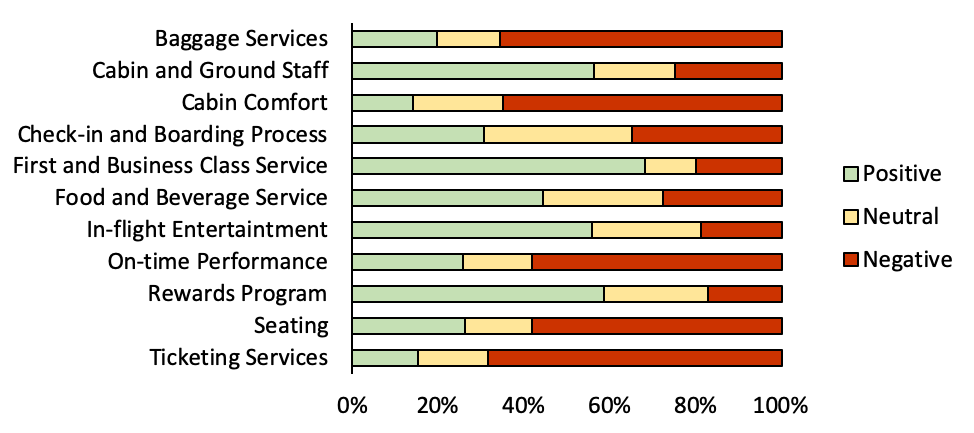}
	\caption{Aspect-based opinion summary for target airline}
	\label{fig:target}
\end{figure}

Figure \ref{fig:target} illustrates the AOS summary for the target airline from the passenger’s perspective. It is evident that “Rewards Program”, “In-flight Entertainment”, “First and Business Class Service”, “Cabin and Ground Staff” and “Food and Beverage” are the aspects that are positively perceived by a majority of the passengers, and thus can be regarded as the target airline’s strengths. On the contrary, most passengers have lamentable  experiences or grievances on the following five aspects of service quality: “Seating”, “On-time Performance”, “Ticketing Services”, “Cabin Comfort”, and “Baggage Services”, and hence are weaknesses. The check-in and boarding process received mixed opinions from the passengers, thereby indicating scope for improvement. 

To ascertain the underlying attributes for customer satisfaction/dissatisfaction in  the target airline, the aspects  identified as its strengths and weaknesses were further explored by analyzing the bi-grams that occur within the review sentences. Specifically, the frequently occurring bi-grams were obtained by analyzing review sentences stratified by aspects. Following  Srinivas and Rajendran \cite{Srinivas2019}, a bi-gram was considered to be frequent if it appears in more than 15\% of the processed reviews. 

\begin{figure}[h]
	\centering
	
	\captionsetup{justification=centering}
	\includegraphics[width=1\linewidth]{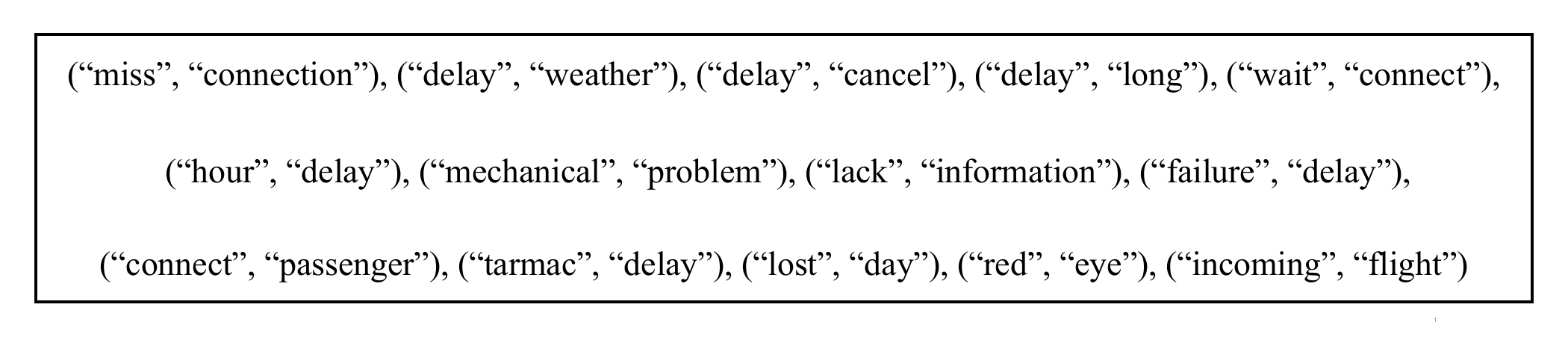}
	\caption{Frequent bi-grams from review sentences expressing negative sentiment towards “On-time Performance” aspect}
	\label{fig:Fig4}
\end{figure}

\begin{figure}[h]
	\centering
	\captionsetup{justification=centering}
	\includegraphics[width=1\linewidth]{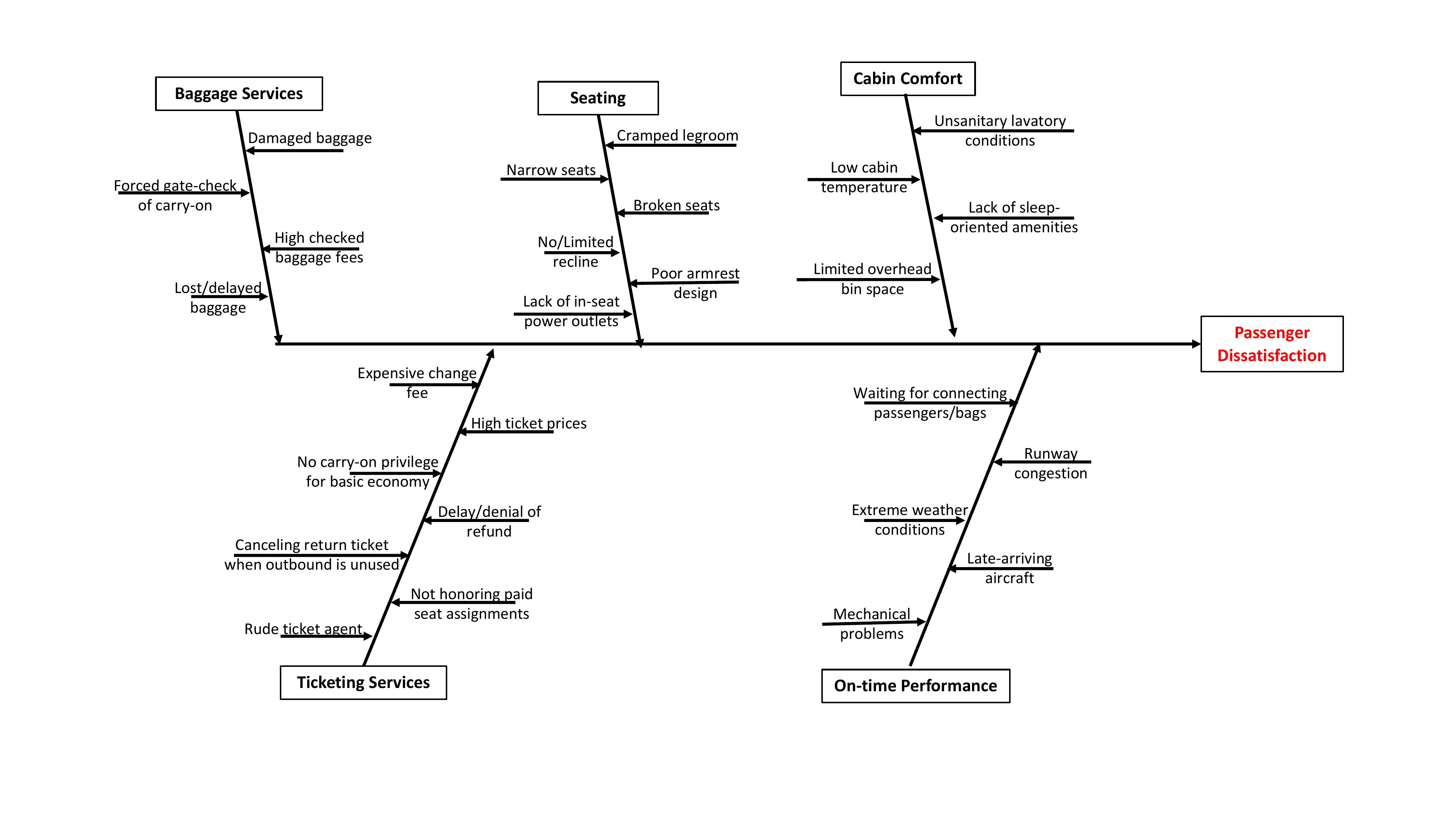}
	\caption{Root cause analysis of topics with overall negative sentiment}
	\label{fig:Fig5}
\end{figure}

Figure \ref{fig:Fig4} provides the most frequently occurring bi-grams from all the review sentences that express a negative opinion towards “On-time Performance”.  It can be observed that the bi-grams such as (“incoming, flight”) and (“wait”, “connect”) are common, and thus we can infer the late-arrival of aircraft to be an underlying cause of poor on-time performance. Likewise, other causes of dissatisfaction related to delays can be identified as mechanical problems, weather conditions, waiting for connecting bags/passengers, and runway congestion. Similarly, it is possible to deduce the reasons for dissatisfaction pertaining to other weaknesses. The result of the root cause analysis for passenger dissatisfaction is illustrated using the cause and effect diagram in Figure \ref{fig:Fig5}, where the primary causes are the aspects identified as weaknesses and the secondary causes are the attributes determined using bi-gram analysis.

Passengers feel the baggage services to be inefficient mainly due to loss, delay or damage of checked baggage. In addition, customers perceive the fee for luggage check-in to be expensive and also criticize the practice of coercing them to check-in their carry-on items at the departure gate due to limited  overhead bin space. For seating, the most common complaints among the passengers are due to the lack of space and reclining seats. Other feedbacks pertaining to seating are on  improper maintenance and poor design. Specifically, passengers repine about the inadvertent pressing of the call button on the armrest and non-functional/missing power outlets. The underlying causes  of cabin discomfort are  unsanitary aircraft lavatories, limited storage space and low cabin temperature. Additionally, the issue of cold temperature is not remedied with blankets. As far as ticketing services is concerned, the prominent reasons for dissatisfaction are the costs associated with purchasing and rescheduling a ticket. Besides, passengers feel that it is inequitable to change pre-selected and prepaid seats without providing any prior notification/reasoning. While the experience with cabin and ground staff is described to be pleasant, their interaction with the ticket agent is perceived to be impertinent. Finally, improper/obscure communication of ticket policies leads to a series of inconveniences to the customer, namely, delay/denial of refund, unprecedented automatic cancellation of return ticket when outbound is unutilized, and unexpected lack of carry-on privilege. 

\begin{figure}[h]
	\centering
	\captionsetup{justification=centering}
	\includegraphics[width=1\linewidth]{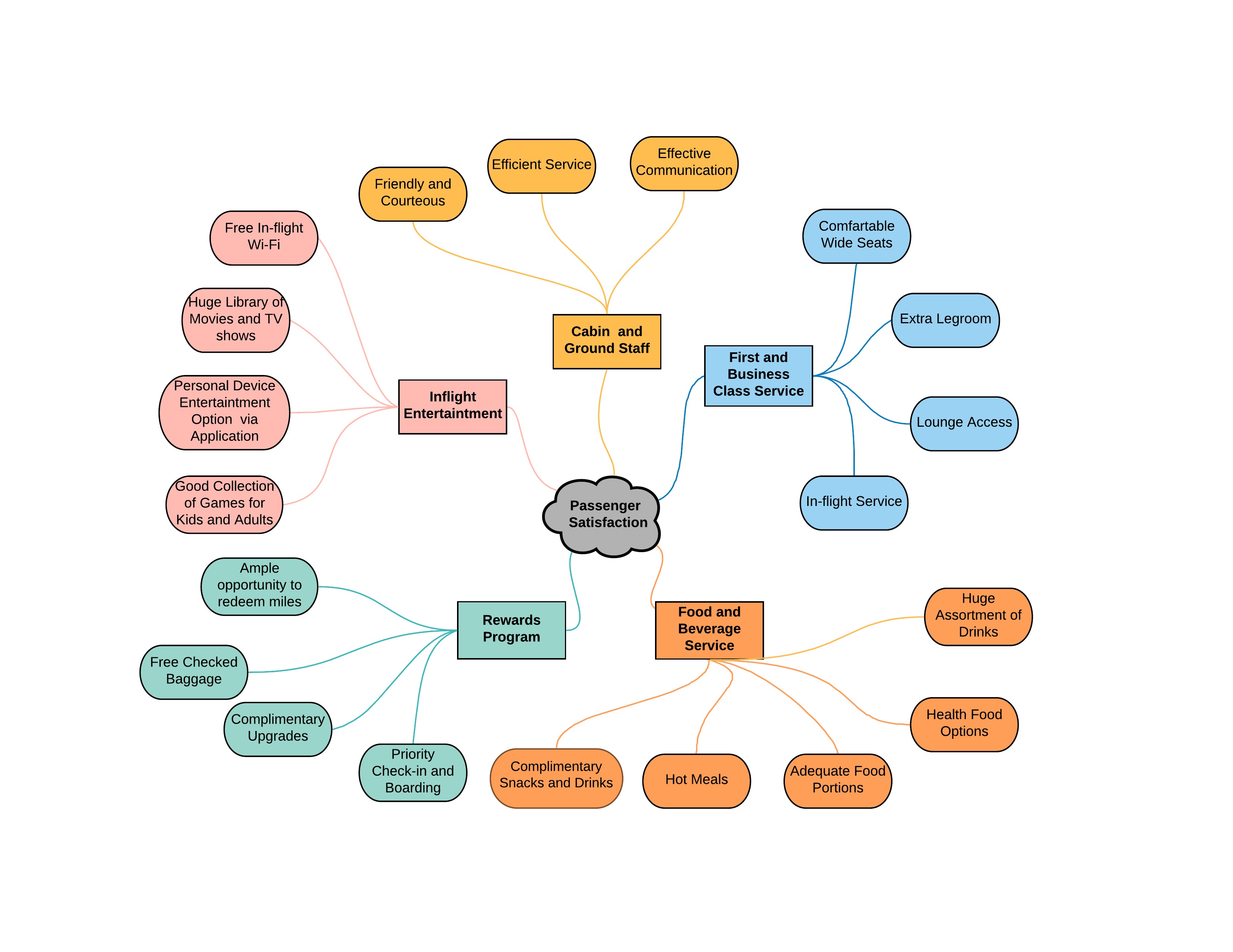}
	\caption{Underlying reasons for passenger satisfaction}
	\label{fig:Fig7}
\end{figure}

Likewise, a similar analysis was conducted on the aspects identified as target airline’s strengths to infer the underlying reasons for passenger contentment, and the corresponding results are summarized in Figure \ref{fig:Fig7}. The cabin and ground staff, one of the important assets of an airline,  is predominantly accoladed  for it’s staff’s  attitude and service delivery. Specifically, the passengers appreciate the staff for promptly responding to customer requests and keeping them well apprised throughout their flight. Regarding in-flight entertainment service, the availability of free wi-fi and  the collection of relevant movies, TV-shows, and music that caters to different demographics are the main reasons for positive reviews. Besides, the flexibility to stream the airline’s media collection on passenger’s personal device (such as tablet and mobile) and age-appropriate alternatives to keep children entertained also play a key role in garnering compliments from the customers. Similarly, for the food and beverage services, the complimentary snacks and soft drinks as well as the huge assortment of in-flight purchase options (such as meals and alcoholic drinks) contribute to passenger satisfaction. Besides, the taste, temperature, and portion of food served exceed their expectations and enhance their in-flight meal experience. As far as business and first-class service is concerned, passengers are delighted about the in-flight service and other amenities such as lounge access. Unlike  most passengers, business and first-class travelers are pleased with their seats as they describe it as comfortable with ample legroom. Finally, for the rewards program, the privileges given on seating, boarding, and baggage allowance are valued by the customers. In addition, travelers are also satisfied with the value of the rewards awarded and the third-party partnerships (e.g., hotels, retailers, car rentals) established by the airline to redeem the points (or miles) accumulated. Thus, the target airline can capitalize on these findings and reinforce them to improve the overall passenger satisfaction.

To gather insights on the competitors, their AOS alongside the target airline's performance was obtained, as shown in Table \ref{tab:Tab4}. The service quality associated with seat comfort appears to be a common problem for all the airlines under consideration, while cabin and ground staff  provide positive customer experience for most of the airlines.   Compared to LCCs, the FSCs are perceived to be better for the rewards program, but worse with respect to ticketing services (such as ticket price, refund, cancellation). However, some carriers appear to be the market leaders for certain aspects. In particular, FSC-1 and LCC-1 are the only carriers that have garnered overall positive reviews for the boarding process and baggage services, respectively. A root-cause analysis of FSC-1 revealed five main check-in and boarding characteristics that were commended by the passengers – (i) seamless, on-time, and quick process, (ii) orderly boarding according to group priority with strategies to  discourage other passengers from blocking the boarding area, (iii) friendly check-in personnel and polite agents (iv) audible announcements for boarding calls, and (v) express  check-in and boarding for  qualified members/ticket holders. Similarly, a bi-gram analysis of review sentences corresponding to LCC-1’s baggage services disclosed their two free checked baggage allowance as the major reason for customer contentment. In addition, passengers also appreciate the timely arrival and the courteous handling of baggage. On the other hand, FSC-2 and LCC-2 are being appreciated for the on-time performance. The frequently occurring bi-grams for these carriers predominantly indicate the early/on-time departure as a key feature for satisfaction. Unlike the target airline’s passengers, these customers rarely complain about mechanical problems, missing connections, or late-arrival of the aircraft. The negative reviews on delay are mostly attributed to extreme weather conditions, which is beyond the airline’s control. This could suggest that the airlines performing well with respect to on-time departures are likely to have (i) robust flight schedules with sufficient buffer times to offset any late arrivals and avoid missed connections, (ii) periodic structural/component maintenance, repair, and overhaul activities to mitigate preventable failures such as fuel tank contamination and worn engine fan blades, and (iii)  sufficient reserves (crew and planes) to handle unavoidable failures such as landing and descent accidents.

\begin{table}[h]
	\centering
	\caption{Root cause analysis of topics with overall negative sentiment}
	\captionsetup{justification=centering}
	\includegraphics[width=1\linewidth]{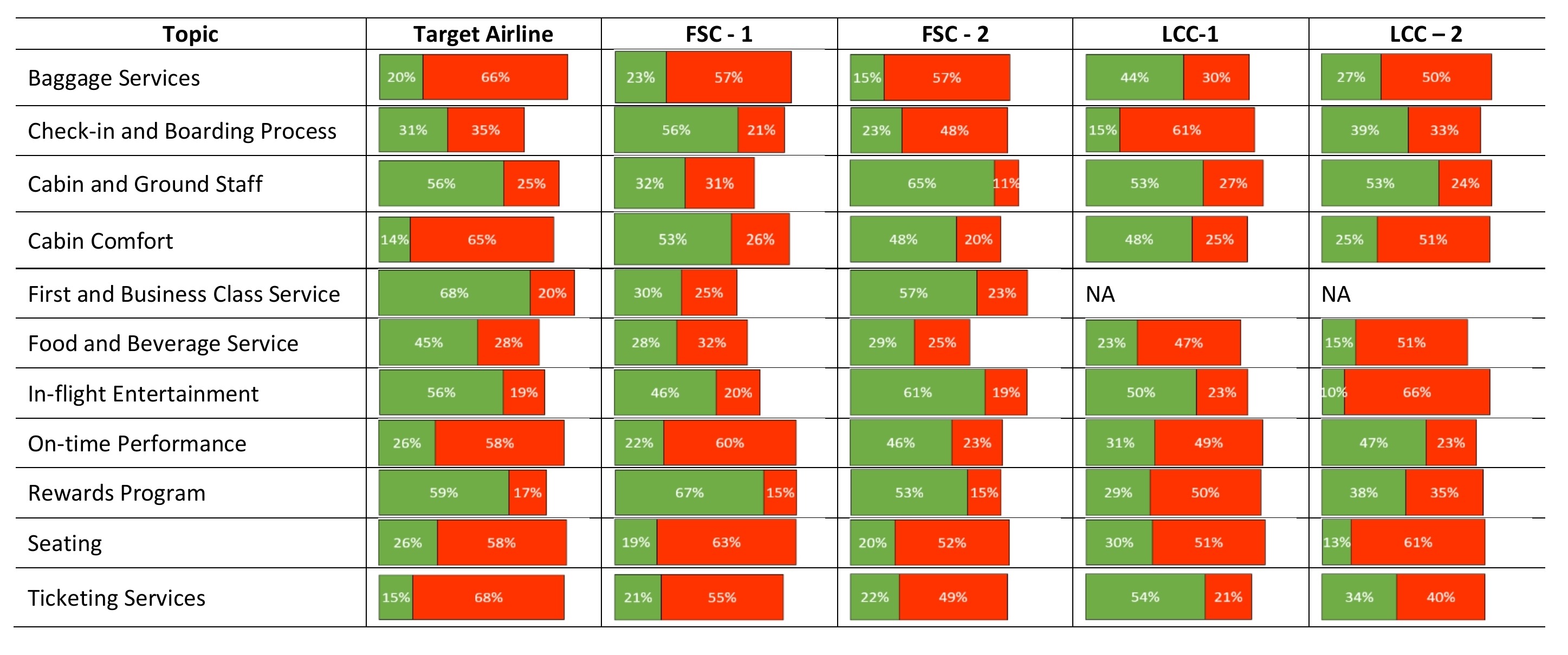}
	\label{tab:Tab4}
\end{table}

In comparison to competitors, the target airline has a competitive edge on food and beverage services but is disadvantaged by  its uncomfortable cabin experience. Moreover, the first and business class service offered by the target airline and FSC-2 is viewed positively by the passengers, while both the LCCs do not offer such services as it is not their primary target market.  Therefore, the target airline can capitalize on the insights obtained from AOS and root cause analysis to strategically focus on the aspects that are perceived poorly by the customers. Specifically, the target airline can learn from the market leaders to improve their business practices on check-in/boarding and baggage services.   In addition, the target airline can adapt the strategies used by FSC-2 and LCC-2 for improving the on-time performance as it is one of the important and most commonly discussed aspects by the passengers.

\section{Discussion}
In this section, we discuss the theoretical, design, and managerial implications of this study. Further, considering the impact of the coronavirus disease 2019 (COVID-19) on air travel and the predictions on more frequent global pandemics in the future \cite{TheIndependent2020}, we also extend the insights obtained from our results to provide implications on preparing flight operations post a disease outbreak (such as SARS, COVID-19). Typically, major disruptions in  airline history have led to industry-wide operational transformations, permanently changing the entire industry  \cite{SITA2020}. For instance, the terrorist attacks on September 11, 2001 lead to heightened security checks at airports to keep air-travel safe \cite{SITA2020}. Similarly, we expect our implications on post-pandemic preparedness to aid in reshaping the future of air travel.

\subsection{Theoretical Implications}
This research reveals staff service, on-time performance, and seat comfort as the most prominent aspects pertaining to airline service quality,  validating prior research \cite{Pakdil2007}. Moreover, most of the dimensions used to measure service quality in airlines concur with the aspects identified by our topic model. For instance, Elliott and Roach \cite{Elliott1993} used timeliness, food and beverage services, luggage transport, check-in process, seat comfort, and in-flight service as dimensions for measuring airline service quality, and these were found among the aspects we extracted as well. Similarly, findings put forth by another recent study \cite{Kim2016}, which recognized seat and cabin comfort as critically important factors, also resurfaced in our analysis. Moreover, Pakdil et al. \cite{Pakdil2007} used the SERVQUAL model to group passengers’ expectations and perceptions into eight different categories that directly match with our extracted aspects or associate as sub groups. In summary, the aspects identified from OCR align with the service quality dimensions provided by traditional methods like SERVQUAL \cite{Parasuraman1988}, and also with the topics extracted in recent studies that utilized text analytics \cite{Korfiatis2019,Sezgen2019,Lucini2020}. 

Many research studies  have shown the significance  of customer service to airline passenger satisfaction, and the current study adds to the depth of knowledge in this area \cite{Mazzeo2003, Sezgen2019}. Sezgen et al. \cite{Sezgen2019} showed that the key driver of satisfaction for a premium class passenger is staff friendliness. This concurs with our results as the carriers that perform well in “First and Business Class Service” also do well in “Cabin and Ground Staff”, and vice versa. Xu et al. \cite{Xu2019} reported that during any service failure or recovery, the primary attribute that decides a passenger’s satisfaction is whether the airline kept them informed by explaining the failure. Our results further support this finding, as the bi-gram analysis indicates lack of information as a factor for dissatisfaction related to on-time performance.

Some findings of this research also contradict the conclusions of existing studies. For instance, Chatterjee \cite{Chatterjee2006} reported that customers are more disposed to come across negative reviews on online platforms as disgruntled consumers are more eager to share their negative experiences. However, the most discussed topic in our study, “Cabin and Ground Staff”, is positively perceived for all airlines. Likewise, Kano et al. \cite{KANO1984} indicated that meeting customer’s basic expectations like on-time performance may not lead to satisfaction/appreciation, but failing in these aspects would lead to dissatisfaction. Though the reviews of the target airline concur to this claim, reviews of FSC-2 and LCC-2 differ as the proportion of positive reviews about on-time performance is almost double the negative ones, indicating that basic expectations receive praise as well. Thus, our results exonerate OCR of its bias attributions, further establishing them as a well-founded source for extracting business intelligence.

Finally, this research also contributes to the literature pertaining to airline review analysis. Specifically, it provides a fresh  perspective on researching aspects of airline service quality by leveraging unsupervised topic models and sentiment analysis methods for discriminative tasks. The proposed ensemble techniques achieves superior classification accuracy for the  topics and sentiments, and also enables a robust AOS for each airline. Such automated hybrid approaches have shown to provide business intelligence in other sectors such as logistics \cite{Rajendran2020} and education \cite{Srinivas2019}. Also, the results of the current study strengthen the support for capitalizing on bi-gram analysis to infer the root cause for customer satisfaction and dissatisfaction in service industry \cite{Rajendran2020}. 

\subsection{Design Implications}
Technological advancements have been reforming the airline industry and have revamped customer experience across all stages of air travel. Airlines can incorporate these technologies into their aircraft design to innovate service and meet customer expectations.

\begin{itemize}
\item \textbf{Leverage Reconfigurable Seating:} As indicated in our analysis, common complaints among passengers of all airlines include cramped legroom, narrow seats, and limited recline. Given that the cabin load factor (i.e., percentage of seats filled with passengers) for domestic flights is typically around 80\%, an astute solution would be to adjust the seat spacing depending on real-time load.  In particular, airlines should consider an overhaul of seating by adopting the concept of smart cabin reconfiguration, where the seating layout for each journey is personalized based on real-time demand, allowing the airlines to use cabin space efficiently \cite{Recaro2017}. For example, the Flex Seat is a sliding seat concept developed jointly by Airbus, Recaro, and THK to give passengers extra legroom and space whenever the flight is not at its full capacity \cite{Recaro2017}.

\item \textbf{Automate Lavatory Disinfecting Process:} Though not mentioned frequently in the reviews, the target airline still has strong negative reviews on lavatory cleanliness. Scheduling more cleaning sessions could ameliorate passenger satisfaction, but it is inconvenient and may not feasible. A better solution is to automate the disinfecting process after each use by retrofitting planes with modern self-cleaning bathrooms developed by Boeing, which uses ultraviolet light (UV) to sanitize all lavatory surfaces in three seconds \cite{Boeing2016}. Deploying such self-cleaning systems, besides improving cabin perception, also helps the cabin staff to focus on other in-flight services.

\item \textbf{Retrofit Redesigned Overhead Baggage Bins}: Inadequate overhead bin space is identified to be a common problem among all airlines under study. Especially, basic economy class flyers are dissatisfied with the lack of carry-on privilege and many others show resentment as their carry-ons are forcefully checked-in at the gate. Incorporating redesigned overhead baggage bins, such as Airbus’s Airspace XL and Boeing’s Space Bins, can help airlines overcome this challenge by accommodating 50-60\% more luggage \cite{AirbusServices2020}. Besides, these incipient designs can be retrofitted to existing fleet of airlines without increasing the aircraft’s weight \cite{AirbusServices2020}

\item \textbf{Provide Personalized Cabin Environment}: As revealed in the analysis, the target airline's passengers perceive the cabin temperature to be uncomfortably cold. Since expectations and perceptions of the cabin environment are subjective, a seat-level control  of the cabin environment is preferred \cite{Grun2013}.  The personalized passenger climate control prototype proposed by ISPACE has proved to provide a more comfortable cabin experience as it empowers  the passenger to control  thermal comfort, humidification and ventilation at seat level \cite{Grun2013}. Further, deploying a personalized displacement ventilation system is also considered efficacious in maintaining optimal thermal comfort \cite{You2019}.

\end{itemize}

\subsection{Managerial Implications}

The results of this study also provide several implications for practical consideration by the airline industry. The following list highlights noteworthy managerial recommendations based on the AOS and bi-grams of the target airline and its competitors.

\begin{itemize}
\item \textbf{Optimize Schedules and Time Buffer:} Nearly 20\% of all reviews comment on the timeliness of the airlines, and the target airline has the worst performance in this category. The competitors (FSC-2 and LCC-2) who performed well in this aspect did not receive any complains on late-arriving aircraft or missed connections. This suggests a scope for amending the  target airline's scheduling activity. Further, this is crucial as a delay that occurs early in the schedule causes a ripple effect that leads to over 70 other flight delays on the that day \cite{Jhunjhunwala2016}. Specifically, the airline should harness the data generated and build analytical models to optimize the schedule and time buffers. 

\item \textbf{Deploy Predictive Maintenance Solutions:} A substantial proportion of reviews also report dissatisfaction due to delays caused by mechanical failure for the target airline, while such complaints are rare for the market leaders (FSC-2 and LCC-2). Employing a predictive maintenance strategy, an artificial intelligence-based approach that monitors equipment's health using sensor data to detect anomalies, can enable an airline to surmount such challenges \cite{Daily2016}. In particular, deploying predictive maintenance solutions allows for early detection of failures and enables timely pre-failure interventions, hence removing any unexpected breakdown, unplanned downtime or flight delays \cite{Daily2016}.

\item \textbf{Introduce Flexible Booking Policy:} While an overwhelming majority of the customers  markedly revere the flexible booking option (i.e., no cancellation charge, no change fee, upfront about costs) of LCC-2,  most passengers of FSCs (target airline, FSC-1 and FSC-2) dread the exorbitant  fee on ticket changes and cancellation, and many more are dissatisfied with the lack or delay of refunds. Therefore, the target airline can adopt policies similar to that of LCC-2 to gain competitive advantage over other FSCs and also tackle the near-monopoly position of LCC-2 with respect to ticketing services. Specifically, the target airline can better prepare to handle flexible booking policy by incorporating the possibility of cancellation and no-shows in their revenue management models and seat booking policies \cite{Yoon2012}. Further, customers are also dismayed  by the insolent  behavior of the ticketing agents. This can be avoided by providing the staffs with better task clarifications, performance-based feedback, and social praise \cite{Reetz2016}. 

\item \textbf{Install Accurate Luggage Tracking Systems:} Aside from forced check-in of carry-on bags, lost, delayed, and damaged baggage are some of the underlying reasons for overall negative sentiments. Though the target airline offers baggage tracking services, customers report the information provided to be incorrect and misleading.  By adopting RFID tags in lieu  of the regular barcode tags, airlines can address this issue in a cost-efficient manner and ensure passenger satisfaction \cite{Rezaei2018}. Further, automation of baggage handling processes at airports, apart from improving on-time performance and lowering costs, will also ameliorate the overall baggage handling and prevent damage \cite{Rezaei2018}. Finally, it is also recommended that airlines provide frequent automated baggage related updates to the passengers’ handheld devices to ensure satisfaction \cite{Rezaei2018}.

\item \textbf{Automate Passenger Identification and Authentication:} Based on reviews, long queues for check-in and boarding agitates the passengers. The use of blockchain technology, coupled with biometrics, can remove the need to present several documents at multiple checkpoints and improve on-time performance \cite{Patel2018}. It does so by creating a biometrics-predicated single token ID that automates passenger identification and authentication, and serves as the boarding pass, passport, and ID for the passenger's travel \cite{Patel2018}. Hence passengers can seamlessly transfer through check-in, baggage drop, immigration, security, and board the aircraft at convenience \cite{Patel2018}.

\end{itemize}

\subsection{ Implications for Post-pandemic Preparedness}

After reaching a near standstill due to the COVID-19, airlines across the world are planning to restart air travel. While preparing the suspended flights for travel is essential, it is also crucial for the airlines to restore passenger confidence by adopting robust measures to prevent the transmission of deadly viruses and bacteria. Further, as people around the world accentuate  on “sanitation” and “social-distancing” during the current pandemic, our research reveals the already existing passenger dissatisfaction with “unsanitary lavatory conditions”, “long queues and waiting times” and “cramped seating”, establishing the challenge ahead. In such a scenario, ascertaining  that the passenger’s voice is heard and addressed will be the key to regaining passenger confidence, the biggest challenge facing the airlines today \cite{SITA2020}.

Incorporating the aforementioned design implications would also prepare the airline to handle a pandemic, where aspects related to ``hygiene/cleanliness", ``non-contact/touchless process" and ``physical distancing" are expected to take precedence. For instance, the recommended UV based self-cleaning lavatory, apart from improving passenger satisfaction, kills 99.99 percent of pathogens \cite{Boeing2016}. Further, research also indicates that UV light can both efficiently and safely inactivate human coronaviruses \cite{Buonanno2020}. Thus, its ability to sanitize, coupled with its touch-free features \cite{Boeing2016} could be a revolutionary approach to deal with infectious diseases. Further, the recommended smart cabin reconfiguration concept can be acclimated to customize the seating layout \cite{Recaro2017} and provide the required distancing depending on the type and the severity of a pandemic. Moreover, apart from maintaining cabin thermal comfort, the recommended personalized ventilation system also prevents the transmission of infections like SARS by reducing the transport of contaminants \cite{You2019}.  

Besides ensuring distancing and sanitary conditions, it is also possible to minimize the number of human-touch points during air travel by adopting the managerial recommendations suggested in this study. For example, the biometrics-predicated single token ID eliminates the need for human intervention \cite{Patel2018} and facilitates contactless boarding, thereby making airports safer during pandemics. Moreover, the automation of baggage handling processes and maintenance systems will reduce worker interventions at the airport, further improving pandemic safety. Finally, during such dubious  times, the suggested ticketing policies and practices can enable airlines to handle cancellations stemming from both government restrictions and passenger precautions. In addition, we have also derived the following insights from our AOS and bi-gram results to enable post-pandemic preparedness.

\begin{itemize}
\item \textbf{Facilitate Informed  Service Withdrawals:} Numerous airlines are fighting against the current pandemic, COVID-19, by slashing services such as in-flight meals and beverages to reduce physical contact \cite{FoxNews2020}. While many of these service withdrawals are crucial for pandemic safety, they ineluctably cause inconvenience to passengers. Our research results will help the airline management anticipate the impact of withdrawing a specific service on customer satisfaction. For example, in-flight sales are a rare discussion in the reviews; with this intelligence, airlines can confidently cut-off this service to ensure pandemic safety. On the other hand, from our bi-gram analysis, it  is discernible that lack of sleep orientated amenities is a significant cause for passenger dissatisfaction. As a result, airlines can inform the passengers beforehand if withdrawing this service is still crucial, thereby changing passenger expectations early and alleviating dissatisfaction.

\item \textbf{Provide Targeted  Service Alterations:}  Few airlines are also combating the pandemic with service alterations \cite{FoxNews2020}. Our research reports the specific aspects of airline services that are vital for customer satisfaction. Instead of eliminating these important services, they can be altered to reduce physical contact. For example, our results indicate that beverage services are crucial to passenger satisfaction. With this astuteness, airline managers can focus on alternatives/alterations before considering the complete withdrawal of this service. A potential alteration is to replace poured beverage service with packaged beverages that can placed on seats before boarding passengers. Similarly, in-flight entertainment is vital for passenger satisfaction; instead of cutting down the entire service due to concerns about handling headphones and contact with high-touch surfaces like LCD screens, passengers can be notified to bring their own devices/accessories on-board or sanitizers could be placed next to such surfaces. 

\item \textbf{Attract and Retain Premium Passengers:} The first and business class passengers constitute close to a third of the total revenue generated in the entire airline industry \cite{Theo2020}. Moreover, in long-haul carriers, this contribution can be as high as seventy prevent \cite{Theo2020}. Attracting this elite class would be vital during a post-pandemic recuperation \cite{Theo2020}. Our research shows that access to lounges, spacious leg-room, comfortable seats, and options to upgrade are essential to this segment. Hence, continuing and improving these services is crucial to attract this lucrative class and drive post-pandemic recovery.

\item \textbf{Maintain Passenger Loyalty:} Downtimes have always been linked with loss in customer loyalty \cite{Bozhikov2012}. Hence, during a pandemic period, airlines must focus on sustaining their loyal passengers. Award programs strive to improve customer loyalty, and our research shows that the mileage program and other frequent flyer benefits like priority seating, extra leg-room, and seat upgrades are vital aspects of the rewards program needed to maintain customer contentment. Ensuring that these passenger miles and frequent flyer benefits are valid when the passengers resume flying post-pandemic will help to maintain passenger loyalty. Further, providing pandemic specific frequent flyer benefits like allocating a seat adjacent to a vacant middle seat can also reinforce passenger loyalty. 

\end{itemize}

\section{Conclusions}
As the airline industry experiences fierce competition and also frequently ranks low in customer satisfaction compared to other service industries, it becomes critical to capture passenger’s expectations and their current perceptions to take measures needed to bridge the service gap and satisfy customers. To automatically gather company-specific and competitor-related passenger perspectives, this research proposed a novel unsupervised text analytics approach that uses online customer reviews (OCR), an economic and abundantly available information source. Three topic models (pLSA, LDA-VI, and LDA-GS) were leveraged to detect vital aspects discussed in OCR, and subsequently an ensemble-assisted approach based on these models were proposed to automatically classify each review sentence into its most prominent topic. Likewise, an ensemble method with three sentiment analysis techniques (AFINN, SentiStrength, and VADER) was developed to predict the opinion pertaining to each review sentence. The sentiment associated with each aspect was then combined to establish the aspect-based opinion summary (AOS), thereby providing an efficient snapshot of customers’ perspectives. Besides, a bi-gram analysis was employed to identify the factors contributing to the current level of customer satisfaction within each aspect, thus facilitating deeper understanding of customer experience and paving way for specific-actionable insights. 

A case study of a US-based airline and its four major competitors  was considered to evaluate the effectiveness of the proposed methodology. The topic models identified 11 essential aspects that were commonly discussed among passengers. The proposed ensemble approaches for sentence-level topic and sentiment classification performed substantially better than the existing individual benchmark models, when tested on a manually-annotated subset of the extracted review sentences. The customer-perceived strengths and weaknesses of each airline were established using AOS, while the common underlying reasons for overall positive/negative perception were revealed by bi-gram analysis. Based on these results, we provided theoretical, design and managerial implications to aid researchers and practitioners with critical decision-making. Finally, from our results, we were also able to derive implications to help airlines restart service post-COVID-19 and prepare better to face future pandemics.

Apart from addressing the literature gap in the effective extraction of intelligence from OCR to improve passenger satisfaction in the airline industry, the proposed innovative approach to capture and analyze the customer's voice also provides numerous directions for future research. First, our findings were based on OCR extracted from a single travel platform for a sample of US-based airlines that have major domestic operations. Future studies can verify the results by collating data from multiple travel platforms and considering airlines specializing in international routes as well as domestic carriers based in other countries. Second, the extracted OCR were not stratified based on comprehensive factors (age, nationality, and educational qualifications of the passenger, flying season, demographics, and cabin segment) and therefore could not provide implications personalized for these attributes. Nevertheless, our proposed methodology can be applied independently on the stratified OCR to obtain personalized decisions for a passenger type in a specific journey, and future works could validate the effectiveness of such an approach. Third, the proposed ensemble approaches were only validated on a subset of airline reviews as it is cost and time intensive to manually annotate large samples for evaluation. Thus, evaluating the proposed unsupervised ensemble techniques on existing annotated data from different domains (e.g., airlines, hotel, manufacturing) is a future research opportunity.  Finally, in addition to affecting customer satisfaction, service quality has also been an important driver of perceived value. As the current study does not explore the interconnection between service quality, satisfaction, and perceived value, future studies  could investigate  the  relationship between the three different factors to gain a better understanding of the predictors of customer satisfaction.

\bibliographystyle{apalike}
\bibliography{AirlineReviews2}

\end{document}